\documentclass{emulateapj}
\usepackage{lscape}

\newcommand{\ee}[1]{\mbox{${} \times 10^{#1}$}}

\newcommand{\h}{$^h$}
\newcommand{\m}{$^m$}
\newcommand{\s}{$^s$}
\newcommand{\degree}{\mbox{$^{\circ}$}}
\newcommand{\am}{\mbox{\arcmin}}
\newcommand{\as}{\mbox{\arcsec}}

\newcommand{\kms}{\mbox{km s$^{-1}$}}
\newcommand{\um}{$\mu$m}

\newcommand{\lsun}{\mbox{L$_\odot$}}
\newcommand{\msun}{\mbox{M$_\odot$}}
\newcommand{\rsun}{\mbox{R$_\odot$}}
\newcommand{\lbol}{\mbox{$L_{bol}$}} 
\newcommand{\lacc}{\mbox{$L_{acc}$}} 
\newcommand{\lint}{\mbox{$L_{int}$}} 
\newcommand{\tbol}{\mbox{$T_{bol}$}} 
\newcommand{\lbolsmm}{$\lbol/$\lsmm}

\newcommand{\ta}{{$T_A^*$}}

\newcommand{\tmb}{\mbox{$T_{\rm mb}$}}

\newcommand{\lsmm}{\mbox{$L_{smm}$}} 

\newcommand{\hh}{\mbox{{\rm H}$_2$}}

\newcommand{\cojone}{$^{12}$CO $J=$ 1$-$0}
\newcommand{\cojtwo}{$^{12}$CO $J=$ 2$-$1}

\newcommand{\cooo}{C$^{18}$O}

\newcommand{\dcop}{DCO$^+$}
\newcommand{\nthp}{N$_2$H$^+$}

\newcommand{\andre}{Andr\'{e}}

\newcommand{\etamb}{$\eta_{\rm mb}$}
\newcommand{\vlsr}{\mbox{$v_{lsr}$}}

\newcommand{\mdotacc}{\mbox{$\dot{M}_{\rm acc}$}}
\newcommand{\taud}{\mbox{$\tau_d$}}

\input{epsf}

\slugcomment{\footnotesize {Accepted by ApJ}}
\begin{document}
\title {\bf The \emph{Spitzer} c2d Survey of Nearby Dense Cores. IX. Discovery of a Very Low Luminosity Object Driving a Molecular Outflow in the Dense Core L673-7}
\author{
Michael M.~Dunham\altaffilmark{1,2},
Neal J.~Evans II\altaffilmark{1},
Tyler L.~Bourke\altaffilmark{3},
Philip C.~Myers\altaffilmark{3},
Tracy L.~Huard\altaffilmark{4},
\& Amelia M.~Stutz\altaffilmark{5,6}
}

\altaffiltext{1}{Department of Astronomy, The University of Texas at Austin, 1 University Station, C1400, Austin, Texas 78712--0259, USA}

\altaffiltext{2}{mdunham@astro.as.utexas.edu}

\altaffiltext{3}{Harvard-Smithsonian Center for Astrophysics, 60 Garden Street, Cambridge, MA 02138, USA}

\altaffiltext{4}{Department of Astronomy, University of Maryland, College Park, MD 20742, USA}

\altaffiltext{5}{Max-Planck-Institut f\"{u}r Astronomie, K\"{o}nigstuhl 17, D-69117 Heidelberg, Germany}

\altaffiltext{6}{Department of Astronomy and Steward Observatory, University of Arizona, 933 North Cherry Avenue, Tucson, AZ 85721, USA}


\begin{abstract}

We present new infrared, submillimeter, and millimeter observations of the dense core L673-7 and report the discovery of a low-luminosity, embedded Class 0 protostar driving a molecular outflow.  L673-7 is seen in absorption against the mid-infrared background in 5.8, 8, and 24 \um\ \emph{Spitzer} images, allowing for a derivation of the column density profile and total enclosed mass of L673-7, independent of dust temperature assumptions.  Estimates of the core mass from these absorption profiles range from $0.2-4.5$ \msun.  Millimeter continuum emission indicates a mass of $\sim$ 2 \msun, both from a direct calculation assuming isothermal dust and from dust radiative transfer models constrained by the millimeter observations.  We use dust radiative transfer models to constrain the internal luminosity of L673-7, defined to be the luminosity of the central source and excluding the luminosity from external heating, to be \lint\ $= 0.01-0.045$ \lsun, with \lint\ $\sim$ 0.04 \lsun\ the most likely value.  L673-7 is thus classified as a very low luminosity object (VeLLO), and is among the lowest luminosity VeLLOs yet studied.  We calculate the kinematic and dynamic properties of the molecular outflow in the standard manner.  From the outflow properties and standard assumptions regarding the driving of outflows, we calculate the time-averaged protostellar mass accretion rate, total protostellar mass accreted, and expected accretion luminosity to be $\langle \dot{M}_{\rm acc} \rangle \geq 1.2 \times 10^{-6} \frac{\mathrm{sin} \, i}{\mathrm{cos}^2 \, i}$ \msun\ yr$^{-1}$, $M_{\rm acc} \geq 0.07 \, \frac{1}{\mathrm{cos} \, i}$ \msun\, and $\lacc \geq 0.36$ \lsun, respectively.  The discrepancy between this calculated \lacc\ and the \lint\ derived from dust radiative transfer models indicates that the current accretion rate is much lower than the average rate over the lifetime of the outflow.  Although the protostar embedded within L673-7 is consistent with currently being substellar, it is unlikely to remain as such given the substantial mass reservoir remaining in the core.

\end{abstract}
\keywords{ISM: individual (L673-7) - stars: formation - stars: low-mass, brown dwarfs}


\section{Introduction}\label{intro}

The \emph{Spitzer Space Telescope} Legacy Project ``From Molecular Cores to Planet Forming Disks'' (c2d; Evans et al.~2003) has completed a $3.6-160$ \um\ imaging survey of nearby, low-mass star-forming regions, including approximately 100 small, dense starless and protostellar cores.  The very first starless core observed by c2d, L1014, was found to harbor an embedded protostar with \lint\ $\sim$ 0.09 \lsun\ (Young et al.~2004), where \lint, the internal luminosity, is the luminosity arising from the central protostar and disk and excludes luminosity from external heating by the interstellar radiation field (ISRF).  The discovery of L1014 led to the definition of very low luminosity objects (VeLLOs; Di Francesco et al.~2007) as objects embedded within dense cores with \lint\ $\leq$ 0.1 \lsun.

L1014 is not the only VeLLO discovered by c2d.  Dunham et al.~(2008) identified 15 candidate VeLLOs in the full c2d sample.  In addition to L1014, three others have been studied in detail: IRAM 04191$+$1522 (hereafter IRAM04191)\footnote{Unlike most VeLLOs, IRAM 04191$+$1522 was previously known to exist based on the detection of warm dust emission and a molecular outflow (\andre\ et al.~1999).} (\lint\ $\sim$ 0.08 \lsun; Dunham et al.~2006), L1521F (\lint\ $\sim$ 0.05 \lsun; Bourke et al.~2006; Terebey et al.~2009), and L328 (\lint\ $\sim$ 0.05 \lsun; Lee et al.~2009).  Other groups have also found objects consistent with the definition of VeLLOs:  Chamaeleon-MMS1, either a VeLLO with \lint\ $\sim$ $0.01-0.02$ \lsun\ or possibly a first hydrostatic core (FHSC) (Belloche et al.~2006), SSTB213 J041757, a candidate embedded proto brown dwarf with \lint\ $\sim$ 0.003 \lsun\ (Barrado et al.~2009), and L1448 IRS2E, a candidate FHSC with \lbol\ $\leq$ 0.1 \lsun\ (Chen et al. 2010).  Assuming spherical mass accretion at the rate predicted by the standard model ($\dot{M}_{acc} \sim 2 \ee{-6}$ \msun\ yr$^{-1}$; Shu, Adams, \& Lizano 1987) onto an object with a typical protostellar radius of $R \sim 3$ \rsun, a protostar located on the stellar/substellar boundary ($M=0.08$ \msun) would have an accretion luminosity, $L_{acc} = GM\dot{M}_{acc} / R$, of $L \sim 1.6$ \lsun.  VeLLOs, with luminosities more than an order of magnitude lower, must either feature mass accretion rates lower than predicted by the standard model, masses below the stellar/substellar boundary, or some combination of the two.

Despite their similar luminosities, the VeLLOs studied to date show very different core and molecular outflow properties.  L1014, IRAM04191, and L1521F all have substantial mass reservoirs remaining in their envelopes ($M_{env} \sim 2-5$ \msun; Young et al.~2004; Dunham et al.~2006; Bourke et al.~2006), whereas L328 appears to only have $M_{env} \sim 0.1$ \msun.  IRAM04191 drives an extended, bright, bipolar molecular outflow detected in single-dish observations (\andre\ et al.~1999).  L1014 drives a weak, compact outflow detected by the Submillimeter Array (Bourke et al.~2005) but not by single-dish observations (Crapsi et al.~2005).  L1521F and L328 both show some hints of outflows in their infrared morphologies (Bourke et al.~2006; Lee et al.~2009), but neither show unambiguous evidence for driving outflows.  Additional VeLLOs must be studied in detail to learn more about the properties of this class of objects.

VeLLOs represent one extreme of the protostellar luminosity distribution, which recent studies have shown is strongly peaked at low luminosities (Dunham et al.~2008; Enoch et al.~2009a; Evans et al.~2009).  These results represent a confirmation and re-statement of the classic ``luminosity problem'' whereby accretion at the standard rate produces accretion luminosities higher than typically observed for embedded protostars (Kenyon et al.~1990).  The luminosity distribution is inconsistent with a constant protostellar mass accretion rate throughout the duration of the protostellar phase and instead suggests that it is highly time variable with extended periods of very low accretion, as first suggested by Kenyon \& Hartmann (1995).  Vorobyov \& Basu (2005a, 2006, 2009) presented simulations showing that material will pile up in a circumstellar disk until the disk becomes gravitationally unstable and dumps its mass onto the protostar in short-lived accretion bursts.  Vorobyov (2009) compared the distribution of mass accretion rates in these simulations to those inferred from the luminosities of protostars in three molecular clouds compiled by Enoch et al.~(2009a) and concluded that their simulations reproduced some of the basic features of the observed distribution of mass accretion rates.  Based on these simulations, Dunham et al.~(2010) presented evolutionary models modified to include a simple treatment of episodic mass accretion in the standard model of the collapse of a singular isothermal sphere (Shu 1977; Shu, Adams, \& Lizano 1987; Young \& Evans 2005) and showed that models with episodic accretion mostly resolved the luminosity problem and improved the match between the observed and predicted protostellar luminosity distributions.

Some direct observational evidence for episodic mass accretion in the embedded phase exists in the form of accretion bursts in Class I sources (e.g., Acosta-Pulido et al.~2007; K\'{o}sp\'{a}l et al.~2007; Fedele et al.~2007), a mismatch between the measured accretion rates onto the disk and protostar of NGC 1333-IRAS 4B (Watson et al.~2007), and episodicity in jets ejected from some protostellar systems (e.g., Lee et al.~2007).  Additionally, evidence for non-steady mass accretion can be found by studying the molecular outflows driven by protostars.  Since outflows are driven by accretion they retain information on the accretion history.  The outflow properties can be used to calculate time-averaged mass accretion rates and thus accretion luminosities that can be directly compared to the protostellar internal luminosities.  Two Class 0 sources studied in this manner, IRAM04191 and L1251A-IRS3, both have accretion luminosities inferred from their outflows much greater than their internal luminosities, possibly indicating lower current accretion rates compared to the time-average (\andre\ et al.~1999; Dunham et al.~2006; Lee et al.~2010).

In this paper we present new infrared, submillimeter, and millimeter observations of the dense core L673-7 and report the discovery of a low-luminosity, embedded protostar driving a molecular outflow.  We provide a brief introduction to L673-7 in \S \ref{sec_l6737}.  A summary of the observations and data reduction is given in \S \ref{observations}.  We present a qualitative discussion of the results in \S \ref{results}, including the discovery of the protostar and the detection of L673-7 in absorption against the mid-infrared background in \S \ref{irs}, a discussion of the evolutionary status of the protostar in \S \ref{evolstatus}, a discussion of the L673-7 core mass in \S \ref{sec_coremass}, and the discovery of the molecular outflow in \S \ref{outflow}.  In \S \ref{darkcore} we use the absorption profile in the mid-infrared to derive the column density profile and total enclosed mass of L673-7.  Dust radiative transfer models used to constrain the protostellar internal luminosity and confirm its status as a VeLLO are discussed in \S \ref{models}.  In \S \ref{outflowmdot}, we calculate the kinematic and dynamic properties of the outflow and use these properties to calculate the time-averaged mass accretion rate onto the protostar and expected accretion luminosity from L673-7.  Finally, \S \ref{discussion} discusses the results and \S \ref{conclusions} presents our conclusions.

\section{L673-7}\label{sec_l6737}

L673-7 is part of the L673 cloud complex from the Lynds (1962) catalog, which lists L673 as an Opacity Class 6 cloud on a relative scale from 1 (least opaque) to 6 (most opaque).  Lee \& Myers (1999) identified 11 cores within this complex, including L673-7, which they found to have a darkness contrast relative to the background of 4 on a relative scale from 1 (lowest contrast) to 4 (highest contrast).  They found that it was not associated with an embedded Young Stellar Object (YSO) based on \emph{IRAS} data and concluded that it was a starless core.  Three related studies searching for infall motions towards starless cores using different dense gas tracers included L673-7 (Lee, Myers, \& Tafalla 1999; Lee, Myers, \& Plume 2004; Sohn et al.~2007); none found any evidence for infall motions in this dense core.  Park, Lee, \& Myers (2004) included L673-7 in their survey of \cojone\ toward starless cores undertaken with the 14m Taeduk Ratio Astronomy Observatory (TRAO) in Korea.  They found that $\sim$60\% of their sample, including L673-7, showed some evidence for line wings that could result from outflows.  They concluded that some of their starless cores may in fact harbor protostars, but higher angular resolution CO maps than the 50\as\ achieved by their data and more sensitive infrared observations than available from \emph{IRAS} were needed to determine which, if any, starless cores were in fact not starless.

The distance to the L673 complex is somewhat uncertain.  Herbig \& Jones (1983) noted, based on optical images, that it is clearly in the foreground of the extensive, large-scale Aquila obscuration, which begins at about 110 pc and continues to about 1 kpc (Weaver 1949), setting an upper limit to the distance to L673.  Herbig \& Jones argue that the most likely distance to L673 is 300 pc based on astrometric methods.  However, a careful reading of their discussion indicates that their results also allow for distances in the range of $\sim 400-600$ pc.  In the following sections we adopt their most likely distance to the L673 complex of 300 pc as the distance to L673-7.  We discuss the effects of larger possible distances in \S \ref{discussion}.

\section{Observations}\label{observations}

\subsection{\emph{Spitzer Space Telescope}}\label{SST}

Observations of L673-7 were obtained with the \emph{Spitzer Space Telescope} (Werner et al.~2004) in three different observing programs:  Program ID (PID) 139 (c2d; Evans et al.~2003), PID 20386 (PI: P.~C.~Myers), and PID 30384 (PI: T.~L.~Bourke).

\subsubsection{c2d}\label{c2d}

\emph{Spitzer} c2d (PID 139) observations of L673-7 were obtained with both the Infrared Array Camera (IRAC; Fazio et al.~2004) and the Multiband Imaging Photometry (MIPS; Rieke et al.~2004).  The IRAC observations were obtained in all four bands, resulting in images at 3.6 (IRAC band 1), 4.5 (IRAC band 2), 5.8 (IRAC band 3), and 8.0 (IRAC band 4) \um.  Two epochs of observations were taken in order to identify and remove asteroids.  The first epoch was observed on 2004 October 7 (AOR 0005152512); the second was observed on 2004 October 8 (AOR 0005153024).  Each epoch of observations consists of two dithered, 12 s images.  The dither size is approximately 10\as, resulting in an observed field-of-view approximately equal to the IRAC 5\am$\times$5\am\ field-of-view.  The total integration time per pixel is 48 s.

The MIPS observations were obtained in the first two photometry bands (24 and 70 \um).  Two epochs were again observed to identify and remove asteroids.  The first epoch was observed on 2004 April 14 (AOR 0009427200); the second was observed on 2004 April 15 (AOR 0009435904).  The MIPS 1 (24 \um) data were obtained in the photometry mode using the small field size and a 3 s exposure time.  A 1 column by 3 row raster map with full-array spacing in both the rows and columns was observed, giving an observed field-of-view of approximately 9\am$\times$18\am\ and a total integration time per pixel of 84 s.  The MIPS 2 (70 \um) data were also obtained in a nearly identical manner except 3 total cycles, rather than 1, were observed per epoch, giving an observed field-of-view of approximately 4.5\am$\times$13\am\ and a total integration time per pixel of $\sim$ 96 s.

The IRAC and MIPS images were processed by the \emph{Spitzer} Science Center (SSC), version S13, to produce basic calibrated data (BCD) images.  The c2d Legacy Team then improved these images to correct for artifacts remaining in the data, produced mosaics, performed source extraction, and compiled a band-merged source catalog.  A complete description of the above steps can be found in Evans et al.~(2007); see also Dunham et al.~(2008) for a summary of the process.

\subsubsection{PID 20386}\label{cores2deeper}

\emph{Spitzer} IRAC and MIPS data were also obtained in PID 20386 (PI: P.~C.~Myers), a program designed to obtain deeper observations of a sub-sample of the c2d cores.  Hereafter we refer to these data as ``cores2deeper'' data.  Two epochs of IRAC observations were obtained in all four bands.  The first epoch was observed on 2005 October 21 (AOR 0014606080); the second was observed on 2005 October 24 (AOR 0014606336).  Each epoch consists of 8 dithered, 30 s images, resulting in an approximately  5\am$\times$5\am\ field-of-view with a total integration time per pixel of 480 s (10 times deeper in total integration time than the c2d observations).  Additional 0.6 s ``high-dynamic range'' (HDR) mode images were obtained in each epoch to enable photometry on bright sources saturated in the longer exposures.

The MIPS observations were again obtained at 24 and 70 \um.  One epoch was observed on 2005 September 27 (AOR 0014614528).  The 24 \um\ data were obtained in the photometry mode using the large field size with 3 cycles and a 10 s exposure time, giving an observed field-of-view of approximately 5.5\am$\times$10.5\am\ and a total integration time per pixel of 300 s ($\sim$ 3.5 times deeper in total integration time than the c2d observations).  The 70 \um\ data were obtained in an identical manner to the 24 \um\ data, giving an observed field-of-view of approximately 3\am$\times$10\am\ and a total integration time per pixel of $\sim$ 190 s ($\sim$ 2 times deeper in total integration time than the c2d observations).

The IRAC and MIPS images were processed by the SSC, version S14 for IRAC and S13 for MIPS, to produce BCD images.  As with the c2d data, these cores2deeper BCD were then post-processed by the c2d pipeline (see references above).

\subsubsection{PID 30384}\label{cores160}

\emph{Spitzer} MIPS data were also obtained in PID 30384 (PI: T.~L.~Bourke), a program designed to obtain MIPS band 3 (160 \um) images of select c2d cores.  Hereafter we refer to these data as ``cores160'' data.  Due to the details of how MIPS operates, we gained additional 24 and 70 \um\ data for free while obtaining the 160 \um\ data.  The data were obtained in a single epoch on 2008 June 23 (AOR 0018160896) in the scan map mode at the medium scan rate with 80\as\ ($\sim$ 1/4 array spacing) return and forward leg cross scan steps.  One cycle of 8 scan legs was obtained, resulting in a total integration time per pixel of 160 s at 24 and 70 \um\ and 16 s at 160 \um.  The 1/4 array spacing between scan legs ensured redundant coverage at 160 \um.  The data were reduced using the MIPS Data Analysis Tool (DAT; Gordon et al. 2005), as described in detail by Stutz et al.~(2007).

\subsection{Caltech Submillimeter Observatory}\label{CSO}

\subsubsection{SHARC-II 350/450 \um\ Continuum}\label{sharc}

Submillimeter continuum observations at 350 and 450 \um\ were obtained with the Submillimeter High Angular Resolution Camera II (SHARC-II; Dowell et al.~2003) at the Caltech Submillimeter Observatory (CSO).  The 350 \um\ observations were obtained in October 2008 and will be fully described in an upcoming publication (M.~M.~Dunham et al.~2010, in preparation).  The 450 \um\ observations were obtained in June 2005 and are described in Wu et al.~(2007).  Details on the observing strategy, data reduction, and source photometry can be found in the above references.  Table \ref{tab_sed}, which lists wavelength, flux density, uncertainty in flux density, aperture diameter, and references for all available photometry on L673-7 (see below), presents flux densities at both wavelengths.  We list the 450 \um\ photometry for completeness; however, due to uncertainties introduced by uncertain calibration at this wavelength and loss of extended emission in the observing mode chosen for these observations (Wu et al.~2007; M.~M.~Dunham et al.~2010, in preparation), we omit this result from all analysis that follows.

\begin{deluxetable}{lcccc}
\tabletypesize{\scriptsize}
\tablewidth{0pt}
\tablecaption{\label{tab_sed}Photometry of L673-7}  
\tablehead{
\colhead{$\lambda$} & \colhead{$S_{\nu}$($\lambda$)} & \colhead{$\sigma$} &
\colhead{Aperture} & \colhead{}\\
\colhead{(\um)} & \colhead{(mJy)} & \colhead{(mJy)} &
\colhead{(arcsec)} & \colhead{Reference\tablenotemark{a}}}
\startdata
3.6  & $<$0.03 & \nodata &\nodata               & 1 \\
4.5  & $<$0.09 & \nodata &\nodata               & 1 \\
5.8  & $<$0.07 & \nodata & \nodata              & 1 \\
8.0  & $<$0.18 & \nodata & \nodata              & 1 \\
24   & 1.30    & 0.43    & 6.0\tablenotemark{b} & 1 \\
70   & 150     & 26      & 18\tablenotemark{b}  & 1 \\
160  & $<$3800 & \nodata & \nodata              & 1 \\
350  & 2700    & 500     & 40                   & 2 \\
450\tablenotemark{c}  & 1100    & 400     & 40                   & 3 \\
1200 & 480     & 26      & 120                  & 4 \\
\enddata\\
\tablenotetext{a}{References - (1) New \emph{Spitzer} observations; (2) M.~M.~Dunham et al.~(2010, in preparation) (3) Wu et al.~(2007); (4) Kauffmann et al.~(2008).}
\tablenotetext{b}{FWHM of \emph{Spitzer} point-spread profile.}
\tablenotetext{c}{Not considered in all of the following analysis due to uncertain calibration and loss of extended emission.}
\end{deluxetable}

\subsubsection{\cojtwo}\label{co}

Observations of \cojtwo\ at 230.537970 GHz were obtained in June and September 2006 at the CSO with the 230 GHz sidecab receiver and 50 MHz AOS backend (Kooi et al.~1998), providing 1024 channels with a resolution of 0.049 MHz per channel.  The average instrument resolution is approximately 2 channels per resolution element, giving an average velocity resolution of 0.13 \kms.  The standard chopper wheel calibration method (Penzias \& Burrus 1973) was used to measure \ta.  The beam FWHM is 32.5\as\ at 230 GHz\footnote{See http://www.submm.caltech.edu/cso/receivers/beams.html}.  The zenith optical depth at 225 GHz (measured via skydips every 10 minutes) varied between approximately $0.05-0.07$ in June 2006 and $0.15-0.18$ in September 2006, and the system temperature varied between approximately $340-400$ in June 2006 and $300-400$ in September 2006.

The main beam efficiency, \etamb, was calculated for the June 2006 data from observations of Jupiter.  No planets were available in September 2006, so \etamb\ for this run was instead calculated by comparing observations of several bright sources to observations of the same sources obtained on previous runs with known efficiencies.  The calculated \etamb\ shows no systematic change between June and September 2006, with an average and standard deviation of 0.77 and 0.08, respectively.  All scans were converted from \ta\ to \tmb\ (mean beam temperature) by dividing by \etamb, and all temperatures given in this paper are in units of \tmb.  The pointing was checked and updated every $1-2$ hours; the standard deviations in pointing were 6\as\ in azimuth and 5\as\ in zenith angle, giving an overall pointing uncertainty of $\sim 8$\as.  Since most of the pointing changes between updates were due to a slow drift with time, 8\as\ is actually an upper limit to the true pointing uncertainty for any given observation.

A nine-point, position-switched map with a spacing of 30\as\ between scans (approximately full-beam spacing) was observed in June 2006.  The map is centered at the position 19\h 21\m 34.9\s\ $+11$\degree 21\am 19.0\as.  A clean off-position of 8500\as\ east (19\h 31\m 01.6\s\ $+11$\degree 21\am 19.0\as) was found using the \cojone\ Galactic Plane Survey (Dame et al.~2001) available through the SkyView Virtual Observatory\footnote{Available at http://skyview.gsfc.nasa.gov/}.  The map was then extended in June and September 2006, always with 30\as\ spacing between scans, following the location of outflow emission seen in individual scans (see \S \ref{outflow}).  A total of 86 distinct positions were observed, giving a total observed area of 21.5 arcmin$^2$.  The average 1$\sigma$ rms in the 86 scans is 0.25 K in the 0.13 \kms\ channels.

\subsection{Other}\label{Other}

Millimeter continuum observations at 1.2 mm of L673-7 were obtained with the Max-Planck Millimetre Bolometer array (MAMBO) at the IRAM 30-m telescope.  Details on the observing strategy and data reduction can be found in Kauffmann et al.~(2008), and Table \ref{tab_sed} lists the 1.2 mm flux density of L673-7 taken from the photometry presented by Kauffmann et al.~(2008).

\section{Results}\label{results}

\subsection{Dark Core and Infrared Point Source}\label{irs}

\begin{figure*}
\epsscale{1.0}
\plottwo{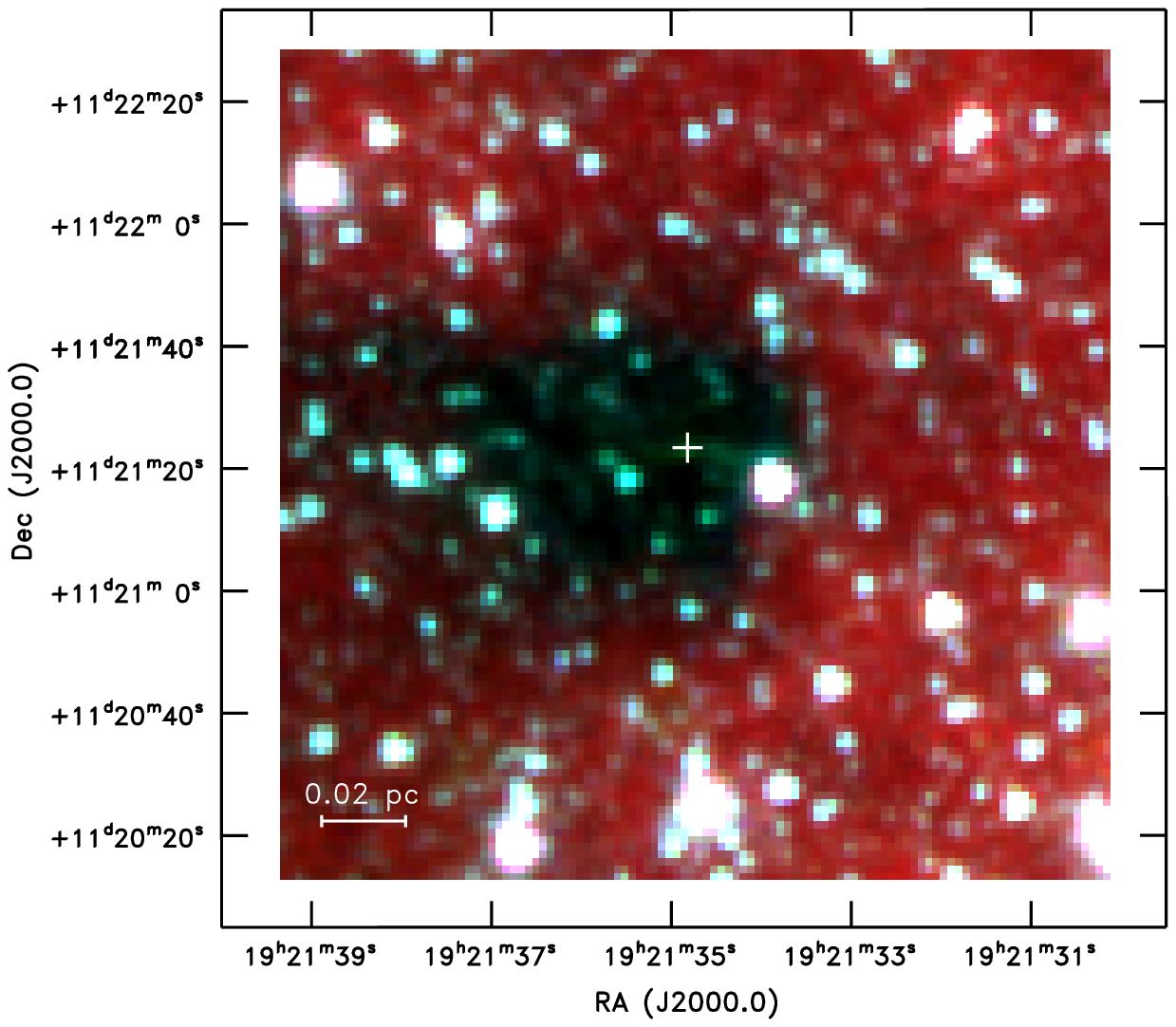}{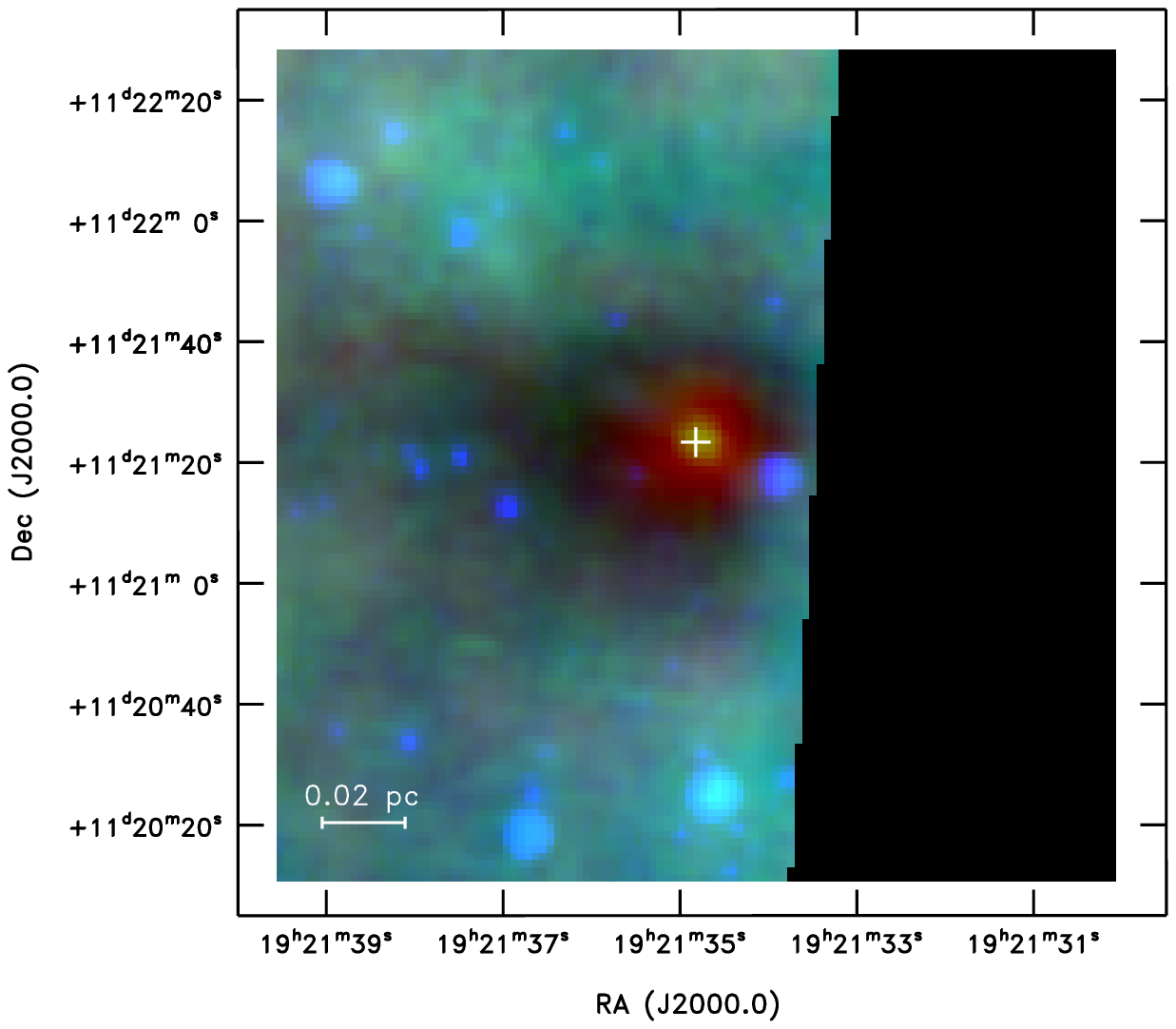}
\caption{\label{fig_3color}\emph{Left:}  Three-color image of L673-7 comprised of the IRAC 1 (3.6 \um; \emph{blue}), IRAC 2 (4.5 \um; \emph{green}), and IRAC 4 (8.0 \um; \emph{red}) c2d images.  The color scales are displayed using a linear stretch in each band, with the minimum and maximum (0.3 and 2.7 MJy sr$^{-1}$ for IRAC 1, 0.15 and 1.8 MJy sr$^{-1}$ for IRAC 2, and 8.75 and 11.5 MJy sr$^{-1}$ for IRAC 4, respectively) chosen to emphasize the dark core seen in absorption against the diffuse background at 8 \um.  The white crosshair marks the position of L673-7-IRS.  \emph{Right:}  Three-color image of L673-7 comprised of the IRAC 4 (8.0 \um; \emph{blue}), MIPS 1 (24 \um; \emph{green}), and MIPS 2 (70 \um; \emph{red}) c2d images.  The color scales are displayed using a linear stretch in each band, with the minimum and maximum (8 and 12 MJy sr$^{-1}$ for IRAC 4, 36 and 38 MJy sr$^{-1}$ for MIPS 1, and 60 and 100 MJy sr$^{-1}$ for MIPS 2, respectively) chosen to emphasize both the point source at 24 and 70 \um\ and the dark core seen in absorption against the 8 and 24 \um\ diffuse backgrounds.  The white crosshair marks the position of L673-7-IRS.  Note that the edge of the 70 \um\ coverage is just to the west of L673-7-IRS, resulting in the black region of no data.}
\end{figure*}

\begin{figure*}
\plotone{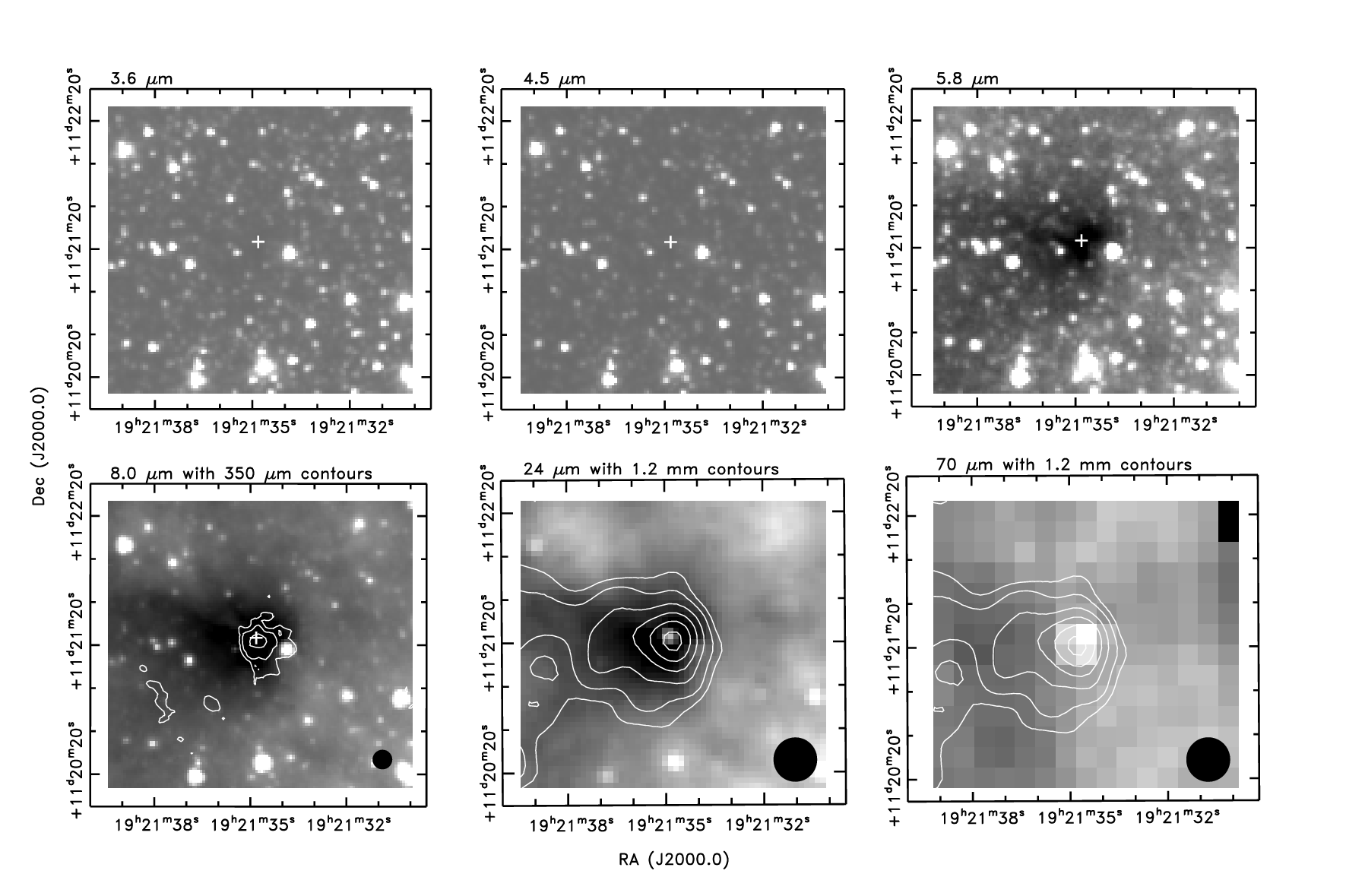}
\caption{\label{fig_cores2deeper}\emph{Spitzer} cores2deeper images of L673-7.  \emph{Top Left:}  3.6 \um\ image displayed in a linear stretch with the minimum and maximum intensities set to $-$3 and 5 MJy sr$^{-1}$, respectively.  The white crosshair marks the position of L673-7-IRS.  \emph{Top Middle:}  4.5 \um\ image displayed in a linear stretch with the minimum and maximum intensities set to $-$3 and 5 MJy sr$^{-1}$, respectively.  The white crosshair marks the position of L673-7-IRS.  \emph{Top Right:}  5.8 \um\ image displayed in a linear stretch with the minimum and maximum intensities set to 2 and 4 MJy sr$^{-1}$, respectively.  The white crosshair marks the position of L673-7-IRS.  \emph{Bottom Left:}  8.0 \um\ image displayed in a linear stretch with the minimum and maximum intensities set to 8.5 and 13 MJy sr$^{-1}$, respectively.  The white crosshair marks the position of L673-7-IRS.  The white contours show the SHARC-II 350 \um\ continuum emission.  The contour levels are [4, 6, 10]$\sigma$, where $1\sigma = 65$ mJy beam$^{-1}$.  The beamsize of the 350 \um\ observations is shown in the bottom right of the panel.  \emph{Bottom Middle:}  24 \um\ image displayed in a linear stretch with the minimum and maximum intensities set to 25.25 and 27 MJy sr$^{-1}$, respectively.  The white contours show the MAMBO 1.2 mm continuum emission.  The contour levels are [5, 7, 10, 13, 16, 19]$\sigma$, where $1\sigma = 1.3$ mJy beam$^{-1}$.  The beamsize of the 1.2 mm observations is shown in the bottom right of the panel.  \emph{Bottom Right:}  70 \um\ image displayed in a linear stretch with the minimum and maximum intensities set to 50 and 71 MJy sr$^{-1}$, respectively.  The contours are the same as in the previous panel.}
\end{figure*}

Figures \ref{fig_3color} and \ref{fig_cores2deeper} show the \emph{Spitzer} images of L673-7.  Figure \ref{fig_3color} shows two three-color images of L673-7: one comprised of IRAC 1 (3.6 \um; \emph{blue}), IRAC 2 (4.5 \um; \emph{green}), and IRAC 4 (8.0 \um; \emph{red}) c2d images, and one comprised of IRAC 4 (8.0 \um; \emph{blue}), MIPS 1 (24 \um; \emph{green}), and MIPS 2 (70 \um; \emph{red}) c2d images.  Figure \ref{fig_cores2deeper} presents individually all of the IRAC and MIPS cores2deeper observations, along with SHARC-II 350 \um\ and MAMBO 1.2 mm continuum emission contours.

Figures \ref{fig_3color} and \ref{fig_cores2deeper} show that the L673-7 core shows up in absorption against the mid-infrared background at 5.8, 8.0, and 24 \um, as confirmed by the 350 \um\ and 1.2 mm continuum emission contours tracing this dark absorption region.  Similar \emph{Spitzer} dark cores have been presented for other low-mass dense cores (e.g., Stutz et al.~2007; Stutz et al.~2009a; Stutz et al.~2009b) and have often been called ``shadows''; here we instead adopt the term ``dark core.''

An infrared point source, SSTc2d J192134.8+112123 (J2000 HHMMSS.s+DDMMSS; hereafter referred to as L673-7-IRS) is visible at 24 and 70 \um\ in Figures \ref{fig_3color} and \ref{fig_cores2deeper}.  L673-7-IRS is not detected at any IRAC wavelength in the c2d band-merged source catalog except IRAC 2.  It does have faint IRAC $1-3$ detections (signal-to-noise ratios (S/N) between $\sim 5-10$) in the cores2deeper catalog.  However, close visual inspection shows no clear evidence for a point source in the IRAC $1-3$ images.  There does appear to be very faint, extended, nebulous emission in these images (not visible in Figures \ref{fig_3color} and \ref{fig_cores2deeper} due to the adopted image scalings) that may have confused the automated point-source extraction routine, but, if present at all, this extended emission is barely above the noise limit and likely arises from a combination of scattered light off the edges of an outflow cavity and emission from molecules excited by the molecular outflow (\S \ref{outflow}; Noriega-Crespo et al.~2004).  For the purposes of assembling a complete spectral energy distribution (SED) to calculate evolutionary indicators (see below) and constrain radiative transfer models (see \S \ref{models}), we treat the IRAC $1-3$ cores2deeper detections as upper limits and list them as such in Table \ref{tab_sed}.  There is no IRAC 4 detection in either the c2d or cores2deeper data and no evidence for emission at the position of L673-7-IRS in the IRAC 4 images.  We calculate a 3$\sigma$ point-source sensitivity of 0.18 mJy for the IRAC 4 cores2deeper image, based on the standard deviation of the sky background, and list this as an upper limit in Table \ref{tab_sed}.

L673-7-IRS is detected at 24 \um\ in the c2d data and listed in the catalog with a flux density of $S_{\nu}^{24 \mu \rm m} = 1.94 \pm 0.27$ mJy (where the uncertaintly includes a 4\% calibration uncertainty [Evans et al.~2007]).  There is no 24 \um\ detection listed in the cores2deeper catalog despite the clear existence of a point source in Figure \ref{fig_cores2deeper}.  We thus performed aperture photometry to calculate the 24 \um\ cores2deeper flux density of L673-7.  We adopted an aperture radius of 7\as, sky annuli of 7\as and 13\as, and an aperture correction of 2.05 as listed on the \emph{Spitzer} Science Center website\footnote{Available at http://ssc.spitzer.caltech.edu/}.  To properly calibrate our aperture photometry with the c2d and cores2deeper catalog PSF photometry, we performed aperture photometry with the same aperture radius, sky annuli, and aperture correction as above on the 50 sources in the cores2deeper catalog with a 24 \um\ detection with S/N $\geq$ 3 and determined an additional correction factor of 0.88 by comparing the results to the catalog flux densities.  With the above parameters, we calculate $S_{\nu}^{24 \mu \rm m} = 1.32 \pm 0.42$ mJy (again including a 4\% calibration uncertainty).

L673-7-IRS is also detected at 24 \um\ in the cores160 data.  We performed aperture photometry using the same method as described above, including an additional correction factor of 0.82 based on comparison of aperture photometry on the cores160 data to cores2deeper catalog photometry.  We calculate $S_{\nu}^{24 \mu \rm m} = 1.10 \pm 0.15$ mJy (where the uncertainty again includes a 4\% calibration uncertainty).  We take a weighted average of the above three results and obtain a final $S_{\nu}^{24 \mu \rm m} = 1.30 \pm 0.43$ mJy, where the uncertainty is the standard deviation of the three separate measurements.  We list this result in Table \ref{tab_sed}.

At 70 \um, the c2d catalog lists a flux density of $S_{\nu}^{70 \mu \rm m} = 148 \pm 35$ mJy (where the uncertaintly includes a 20\% calibration uncertainty [Evans et al.~2007]).  Automatic 70 \um\ source extraction was not performed on the cores2deeper data for technical reasons, and no source extraction was performed on the cores160 data.  Performing aperture photometry using a 16\as\ radius aperture, sky annuli of 18\as\ and 39\as, and an aperture correction of 2.07 (as listed on the SSC website), we calculate $S_{\nu}^{70 \mu \rm m} = 129 \pm 30$ mJy for the cores2deeper data and $S_{\nu}^{70 \mu \rm m} = 181 \pm 37$ mJy for the cores160 data (again including 20\% calibration uncertainties).  No additional corrections are included because there are no other bright 70 \um\ sources in the field for which such corrections can be obtained.  A weighted average of the above values gives a final $S_{\nu}^{70 \mu \rm m} = 150 \pm 26$ mJy, as listed in Table \ref{tab_sed}.

\begin{figure}
\plotone{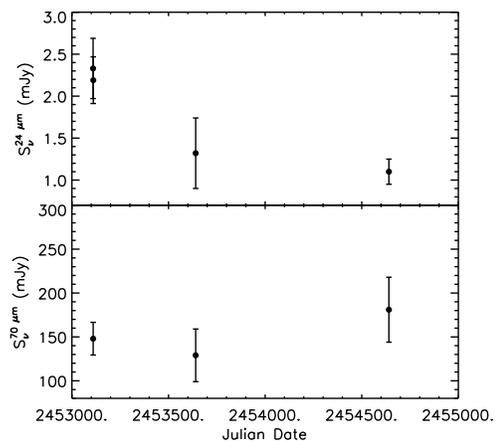}
\caption{\label{fig_variability}$S_{\nu}^{24 \mu \rm m}$ \emph{(top)} and $S_{\nu}^{70 \mu \rm m}$ \emph{(bottom)} vs. time for the multiple epochs of MIPS observations spanning from April 2004 through June 2008.}
\end{figure}

Figure \ref{fig_variability} plots $S_{\nu}^{24 \mu \rm m}$ and $S_{\nu}^{70 \mu \rm m}$ versus Julian Date for the multiple epochs of observations to search for evidence of variability.  There is some evidence for a systematic decrease in $S_{\nu}^{24 \mu \rm m}$ versus time over the four years spanning from April 2004 through June 2008.  However, given the sparse sampling (only four epochs, two of which are separated by only one day) and uncertainties in additional photometry correction factors derived by comparing catalog and aperture photometry flux densities, the case for this decrease remains ambiguous.  There is no evidence for variability at 70 \um\ within the uncertainties.

\begin{figure}
\plotone{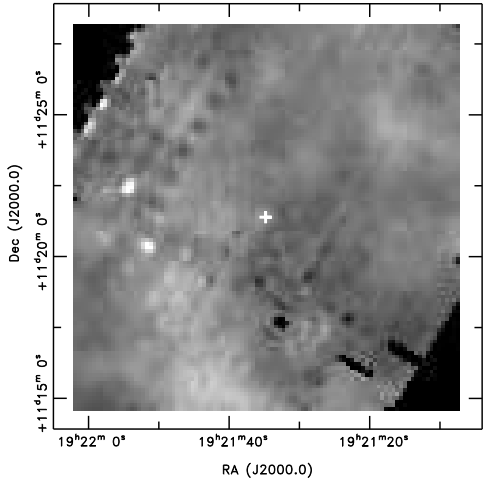}
\caption{\label{fig_mips160}\emph{Spitzer} 160 \um\ cores160 image, displayed in a linear stretch with the minimum and maximum intensities set to $-$75 and $+$75 MJy sr$^{-1}$, respectively.  The white crosshair marks the position of L673-7-IRS.}
\end{figure}

Figure \ref{fig_mips160} shows the 160 \um\ image of L673-7 obtained in the cores160 program.  There is no clear detection associated with L673-7-IRS.  Using the IDL Astronomy User's Library\footnote{Available at http://idlastro.gsfc.nasa.gov/} program sky.pro, which calculates the sky background and variation in sky background using a procedure that iteratively eliminates outlier pixels, we calculate a sky background and variation in this background of -8.0 MJy sr$^{-1}$ and 23.3 MJy sr$^{-1}$, respectively.

Approximating the 160 \um\ \emph{Spitzer} PSF with a Gaussian with $\theta_{\rm FWHM}$ equal to the diffraction limited resolution of $\sim 40$\as, this background intensity is then converted into units of mJy beam $^{-1}$ as follows:
\begin{equation}
\sigma_{\rm sky} \, \rm{(mJy} \, \rm{beam)}^{-1} = 1 \times 10^9  \, \left (\frac{\pi \, \theta_{\rm FWHM}^2}{4 \, ln 2}\right ) \, \sigma_{\rm sky} \, \rm{(MJy} \, \rm{sr)}^{-1} \, ,
\end{equation}
This gives the 1$\sigma$ point source sensitivity since the total flux density of a point source is equivalent to its flux density per beam.  Substituting in the above value of $\sigma_{\rm sky} = 23.3$ MJy sr$^{-1}$ and multiplying by 3 gives a 3$\sigma$ point source sensitivity of 3000 mJy, which we list in Table \ref{tab_sed}.  We consider this to be a conservative upper limit since it encompasses large-scale sky variation over the map on top of the true local variation at the position of L673-7.

Before moving on, we note here that there are 7 sources in the c2d source catalog for L673-7 classified as candidate young stellar objects based on the selection criteria developed by Harvey et al.~(2007).  L673-7-IRS is not among these 7 sources because it is not detected at all \emph{Spitzer} wavelengths between $3.6-24$ \um, a requirement imposed by the Harvey et al.~selection criteria.  These 7 sources have $2-24$ \um\ spectral indices (see Evans et al.~[2007] for details on the calculation of these indices) ranging from $-$2.52 to $-1.82$ and are thus classified as Class III objects (Greene et al.~1994; Evans et al.~2009).  None are clearly associated with L673-7 itself; they are scattered about the $\sim$ 5\am $\times$ 5\am\ field-of-view, with the closest of the 7 located 100\as\ from L673-7-IRS (a projected separation of 30,000 AU at 300 pc).  All 7 sources are classified as AGB stars according to the color and magnitude criteria developed by Robitaille et al.~(2008) to perform a rough separation of YSO and AGB stars.  Finally, L673-7 is located close to the galactic plane in the direction of the inner galaxy ($l \sim 46.5$\degree, $b \sim -1.5$\degree), and Robitaille et al.~(2008) found that the surface density of AGB stars increases exponentially with both decreasing galactic longitude and decreasing galactic latitude.  Thus, we conclude that these 7 sources classified as candidate YSOs are in fact likely background AGB stars.

\subsection{Evolutionary Status of L673-7-IRS}\label{evolstatus}

\begin{figure}
\plotone{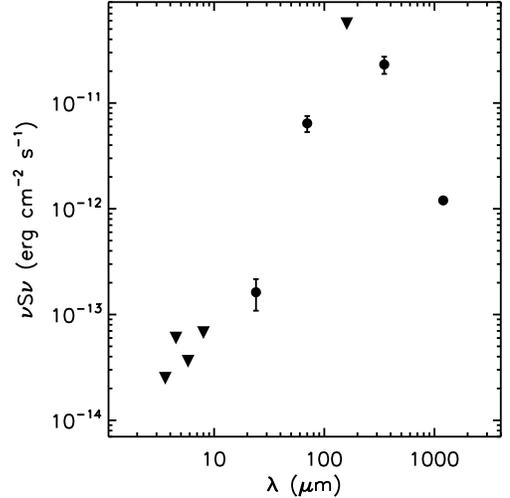}
\caption{\label{fig_sed}Observed SED of L673-7.  Detections are plotted as filled circles with error bars and upper limits are plotted as filled inverted triangles.  See the text for details on photometry at each wavelength.}
\end{figure}

L673-7-IRS features a rising SED in the infrared (Table \ref{tab_sed}), it is located within a dark absorption core in the mid-infrared and associated with submillimeter and millimeter continuum emission, and it drives a molecular outflow (see \S \ref{outflow}).  Thus, L673-7-IRS is an embedded protostar, based on the criteria given by Dunham et al.~(2008).  Figure \ref{fig_sed} plots the observed SED of L673-7, including \emph{Spitzer} MIPS detections at 24 and 70 \um, the CSO SHARC-II detections at 350 \um, and a MAMBO 1.2 mm detection.  Also shown are \emph{Spitzer} IRAC $3.6-8.0$ \um\ and MIPS 160 \um\ upper limits .  Based on the 24, 70, 350, and 1200 \um\ detections we calculate \lbol\ $=0.18$ \lsun, \tbol\ $=16$ K, and \lbolsmm\ $=7$ (where \lsmm\ is defined to be the luminosity longward of 350 \um).  The integrals are evaluated using the trapezoid rule (see Dunham et al.~[2008] for details and a discussion of errors introduced by integrating over finite, incompletely sampled SEDs).  Based on both the calculated \tbol\ and \lbolsmm, L673-7 is classified as a Class 0 object (\andre\ et al.~1993; Chen et al.~1995).

The calculated \lbol\ is likely a lower limit to the true bolometric luminosity since the SED is not fully sampled near its peak and the photometry at $\lambda \geq 100$ \um\ is performed in apertures likely to be smaller than the total source extent.  Furthermore, the calculated \lbol\ is not a good indicator of the total \lint\ since most of the bolometric luminosity likely arises from external heating by the interstellar radiation field (ISRF), which can add up to a few tenths of a solar luminosity to the bolometric luminosity of an embedded, low-mass protostar (e.g., Evans et al.~2001).  We can obtain an estimate of \lint\ from the relation between \lint\ and 70 \um\ flux found by Dunham et al.~(2008).  Applying their relation gives \lint\ $=0.04$ \lsun.  This estimate is good only to within about a factor of two.  Radiative transfer models are required to better constrain both \lbol\ and \lint\ and are presented in \S \ref{models}.

\subsection{Core Mass}\label{sec_coremass}

An estimate of the mass of the L673-7 core can be obtained from the 1.2 mm MAMBO data.  Assuming that the 1.2 mm continuum emission is optically thin and the core is isothermal, the total mass of the core is given as

\begin{equation}\label{eq_dustmass}
M = 100 \frac{d^2 S_{\nu}}{B_{\nu}(T_D) \kappa_{\nu}} \quad ,
\end{equation}
where $S_{\nu}$ is the total flux density, $B_{\nu}(T_D)$ is the Planck function at the isothermal dust temperature $T_D$, $\kappa_{\nu}$ is the dust opacity, $d = 300$ pc, and the factor of 100 is the assumed gas-to-dust ratio.  For consistency with later sections (see \S \ref{darkcore} and \ref{models}), we adopt $\kappa_{\nu}^{1.2 \rm mm} = 1.02$ cm$^2$ gm$^{-1}$, the dust opacity of Ossenkopf \& Henning (1994) appropriate for thin ice mantles after $10^5$ yr of coagulation at a gas density of $10^6$ cm$^{-3}$ (OH5 dust).  With the flux density listed in Table \ref{tab_sed} and an assumed $T_D = 10.5$ K (the derived kinetic temperature, $T_{kin}$, of the core from NH$_3$ observations; T.~L.~Bourke et al.~(2010), in preparation), Equation \ref{eq_dustmass} gives an estimated core mass of 1.9 \msun.  We list this result in Table \ref{tab_coremass}.

\begin{deluxetable*}{lcc}
\tabletypesize{\scriptsize}
\tablewidth{0pt}
\tablecaption{\label{tab_coremass}L673-7 Core Mass Estimates}  
\tablehead{
\colhead{}          & \colhead{Aperture Radius} & \colhead{Mass}     \\
\colhead{Source}    & \colhead{(arcsec)}        & \colhead{(\msun)}}
\startdata
1.2 mm flux density & 60                        & 1.9                \\
8 \um\ dark core    & 70                        & $0.6-1.9$          \\
24 \um\ dark core   & 70                        & $0.2-4.5$          \\
Dust Model          & 83.3                      & 2.25               \\
\enddata\\
\end{deluxetable*}

\subsection{Molecular Outflow}\label{outflow}

\begin{figure}
\plotone{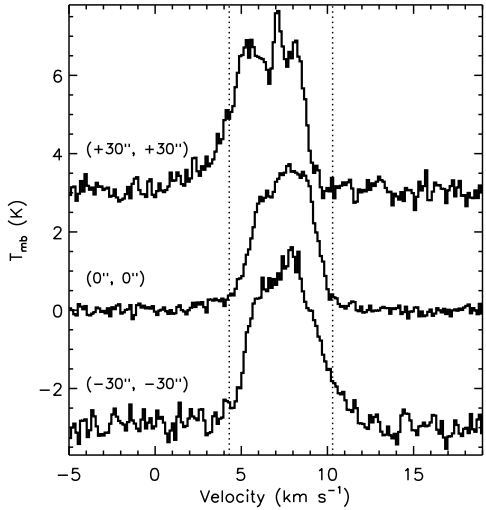}
\caption{\label{fig_outflowspectra}\cojtwo\ spectra at three positions observed within the initial nine-point map:  (0\as, 0\as), (+30\as, +30\as), and ($-$30\as, $-$30\as).  The spectrum at (+30\as, +30\as) is offset by $+$3 K.  It clearly shows a wing of blue-shifted emission relative to the spectrum at (0\as, 0\as).  The spectrum at (-30\as, -30\as) is offset by $-$3 K.  It clearly shows a wing of red-shifted emission relative to the spectrum at (0\as, 0\as).  The dotted horizontal lines indicate the zero-levels of each spectrum, and the dotted vertical lines mark $\pm$ 3 km s$^{-1}$ relative to the core systemic velocity of 7.3 km s$^{-1}$.}
\end{figure}

As stated in \S \ref{co}, we initially observed a nine-point map with 30\as\ spacing centered on L673-7-IRS\footnote{The position of L673-7-IRS and the center of the \cojtwo\ map actually differ by 4.6\as\ because the position used for the CO observations was derived from an early version of the c2d data processing pipeline.  Since this offset is smaller than the pointing uncertainty in the \cojtwo\ map (8\as) and significantly smaller than both the spacing between scans and the beamsize (30\as\ and 32.5\as, respectively), the slight position offset between the two is negligible.}.  Evidence for a molecular outflow was seen in this initial map in the form of a wing of blueshifted emission to the northeast and a wing of redshifted emission to the southwest.  This evidence is shown in Figure \ref{fig_outflowspectra}, which shows the \cojtwo\ spectra at three positions within the initial nine-point map.  We then extended the map to follow this apparent northeast-southwest outflow axis.

\begin{figure}
\plotone{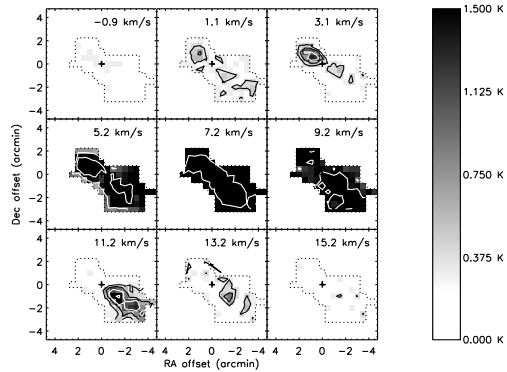}
\caption{\label{fig_channelmap}\cojtwo\ channel maps of the L673-7 outflow with a velocity resolution of 2 \kms.  The grayscale shows main-beam temperature in K, with the mapping between grayscale and value of \tmb\ given in the legend on the right.  The contours are drawn at 2, 5, 10, 20, 35, and 50$\sigma$, where the average 1$\sigma$ rms is 0.095 K in the 2 \kms\ channels.  The lowest two contours are drawn with a black line; all others are drawn in white.  The cross in each panel marks the position of the protostar, and the dotted line encloses the total area mapped.}
\end{figure}

Figure \ref{fig_channelmap} shows channel maps of the CO data at a velocity resolution of 2 \kms.  The core systemic velocity is taken to be \vlsr\ $=7.3$ \kms\ from observations of \cooo\ $J=$ 2$-$1, \nthp\ $J=$ 3$-$2, and \dcop\ $J=$ 3$-$2 emission (J.~H.~Chen et al.~2010, in preparation).  The central three channels are dominated by ambient emission, but an outflow is clearly detected in the higher velocity channels.  Most of the outflow emission is found at $|v-v_{core}| \leq 4$ \kms\ (where $v_{core}$ is the core systemic velocity), with no outflow emission detected beyond $|v-v_{core}| \sim 8$ \kms.

\begin{figure}
\plotone{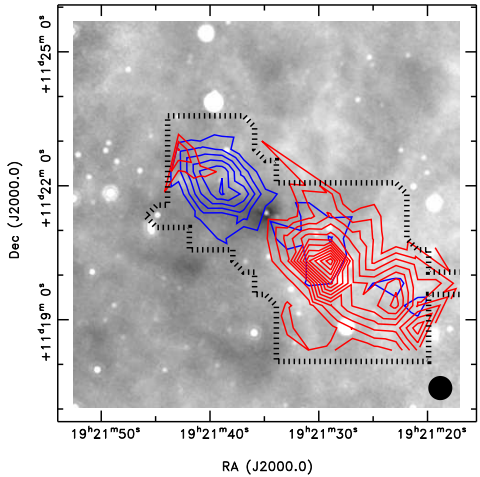}
\caption{\label{fig_outflowcontours}\cojtwo\ integrated intensity contours of L673-7 showing blueshifted and red-shifted emission, overlaid on a 24 \um\ greyscale image centered on L673-7-IRS.  The blue contours are integrated from -2.0 to 4.0 \kms, while the red contours are integrated from 10.0 to 16.0 \kms.  The contours start at 0.5 K \kms\ and increase by 0.5 K \kms.  The dotted line encloses the total area mapped.}
\end{figure}

Figure \ref{fig_outflowcontours} shows integrated redshifted and blueshifted emission.  There is no obvious outflow emission at the map center, but there is one prominent blueshifted peak to the northeast and a corresponding redshifted peak to the southwest.  The total extent of each peak is $\sim$150\as\ from the center, and the position angle of the axis connecting the blue- and redshifted emission is approximately 55\degree\ (measured east [counter-clockwise] from north).  The fact that this outflow shows a clear bipolar morphology with mostly, but not entirely, distinct blue- and redshifted lobes and no outflow emission at the center suggests a moderate, but slightly edge-on, inclination ($i\sim 45-70$\degree; between cases 2 and 3 of Cabrit \& Bertout [1986]).  A more precise estimate of the inclination would require knowledge of the outflow cone opening angle, but extreme pole-on and edge-on inclinations can be ruled out (the former by the bipolar morphology and the latter by the mostly distinct red and blue lobes).  There are additional redshifted peaks along the same axis located farther to the southwest, extending to the edge of the map, but the presence of corresponding blueshifted peaks is unknown since the map does not extend far enough to the northeast.  The full extent of the molecular outflow driven by L673-7-IRS thus remains unknown.

\section{Dark Core: Column Density Profile and Enclosed Mass}\label{darkcore}

Figures \ref{fig_3color} and \ref{fig_cores2deeper} clearly show a dark region in the 8 and 24 \um\ images that corresponds with the 350 \um\ and 1.2 mm dust continuum emission.  These dark regions result from absorption of the mid-infrared background by the dense core L673-7 and can be used to study the column density profile and total enclosed mass of the core, as has been done for prestellar cores with both \emph{Infrared Space Observatory (ISO)} data (e.g., Bacmann et al.~2000) and, more recently, \emph{Spitzer} data (e.g., Stutz et al.~2007, 2009a, 2009b).  The following analysis is based on the method of Stutz et al., and uses the cores2deeper data.

The observed intensity at a given frequency, $I_{\nu}$, is given by 

\begin{equation}
I_{\nu} = I_{\nu}^{o} e^{-\tau_{\nu}} + I_{\nu}^{fg} \quad ,
\end{equation}
where $I_{\nu}^{o}$ is the background intensity, $I_{\nu}^{fg}$ is the foreground intensity, and $\tau_{\nu}$ is the total optical depth.  Knowledge of $I_{\nu}^{fg}$ is important for deriving an accurate intensity profile.  Underestimating $I_{\nu}^{fg}$ will cause one to underestimate the optical depth and corresponding column density, while overestimating $I_{\nu}^{fg}$ will cause one to overestimate the optical depth and corresponding column density.  Stutz et al.~(2007) estimated $I_{\nu}^{fg}$ from the darkest pixels in their map.  However, the darkest pixels in both the 8 and 24 \um\ maps of L673-7 are found at the center of the dark core.  Thus, since we have no direct way of measuring $I_{\nu}^{fg}$, we calculate lower and upper limits to the true absorption profiles by assuming either that there is no foreground emission ($I_{\nu}^{fg} = 0$ MJy sr$^{-1}$) or that all of the emission in the darkest pixel is foreground emission (i.e., that the background emission is completely absorbed in the darkest pixel).  In the latter case the assumed $I_{\nu}^{fg}$ is subtracted from the image before the analysis below is conducted.

\begin{figure}
\plotone{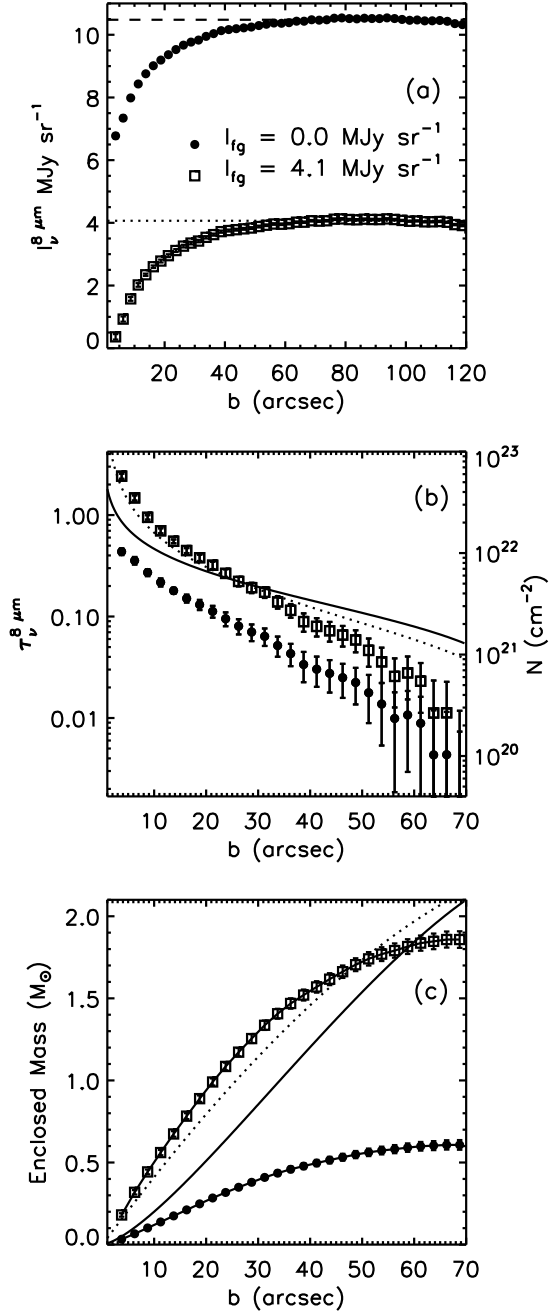}
\caption{\label{fig_darkcore8}Analysis of the 8 \um\ dark core.  In all three panels the filled circles with error bars show the results from assuming no foreground emission ($I_{\nu}^{fg} = 0$ MJy sr$^{-1}$) while the open squares with error bars show the results from assuming that all of the emission in the darkest pixel is foreground emission.  \emph{(a):}  Intensity profile, calculated by averaging over concentric annuli, each with radii of 2.5\as, centered at the position of L673-7-IRS.  The dashed and dotted lines show the background intensities obtained as described in the text for the assumptions of zero and maximum foreground intensity, respectively.  \emph{(b):}  Optical depth profile, calculated from Equation \ref{eq_tau} (\emph{left axis}), and column density profile, calculated from Equation \ref{eq_N} (\emph{right axis}).  The solid and dotted lines show the column density profiles for the best-fit $p=1.5$ and $p=2.0$ models, respectively (see \S \ref{models}).  \emph{(c):}  Total enclosed mass, calculated from Equation \ref{eq_mass}.   The solid and dotted lines show the total enclosed mass for the best-fit $p=1.5$ and $p=2.0$ models, respectively (see \S \ref{models}).} 
\end{figure}

\begin{figure}
\plotone{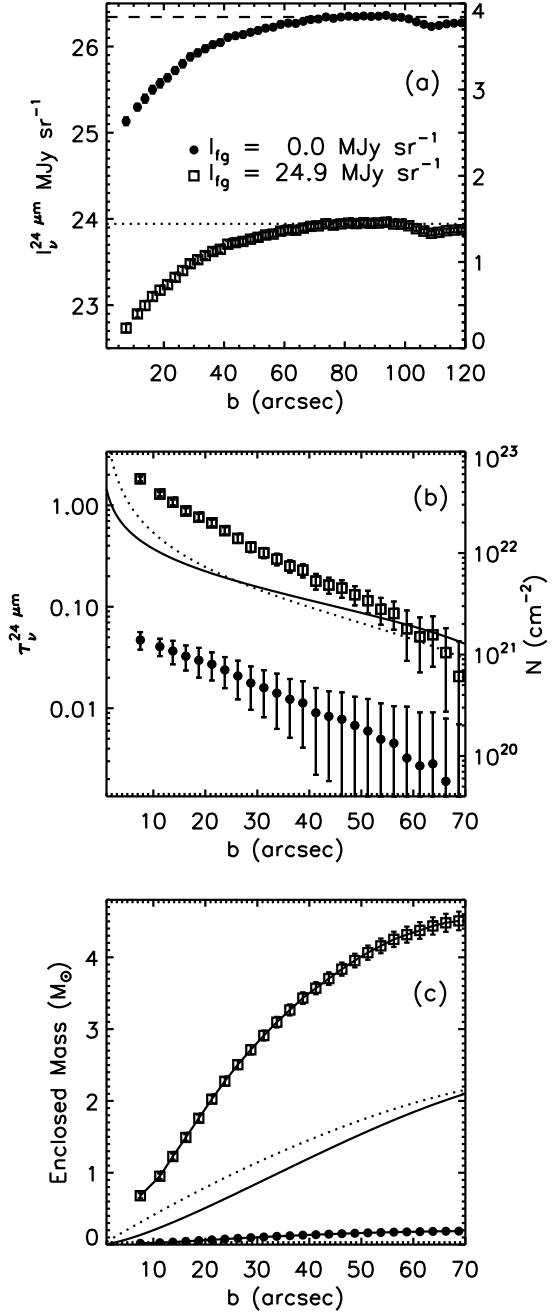}
\caption{\label{fig_darkcore24}Same as Figure \ref{fig_darkcore8}, except now for the 24 \um\ dark core.  In all three panels the filled circles with error bars show the results from assuming no foreground emission ($I_{\nu}^{fg} = 0$ MJy sr$^{-1}$) while the open squares with error bars show the results from assuming that all of the emission in the darkest pixel is foreground emission.  \emph{(a):}  Intensity profile, calculated by averaging over concentric annuli, each with radii of 5\as, centered at the position of L673-7-IRS.  The left y-axis scale corresponds to the $I_{\nu}^{fg} = 0$ MJy sr$^{-1}$ intensity profile while the right y-axis scale corresponds to the $I_{\nu}^{fg} = 24.9$ MJy sr$^{-1}$ intensity profile.  The dashed and dotted lines show the background intensities obtained as described in the text for the assumptions of zero and maximum foreground intensity, respectively.  \emph{(b):}  Optical depth profile, calculated from Equation \ref{eq_tau} (\emph{left axis}), and column density profile, calculated from Equation \ref{eq_N} (\emph{right axis}).  The solid and dotted lines show the column density profiles for the best-fit $p=1.5$ and $p=2.0$ models, respectively (see \S \ref{models}).  \emph{(c):}  Total enclosed mass, calculated from Equation \ref{eq_mass}.   The solid and dotted lines show the total enclosed mass for the best-fit $p=1.5$ and $p=2.0$ models, respectively (see \S \ref{models}).}
\end{figure}

Figure \ref{fig_darkcore8}a shows the 8 \um\ intensity profiles under the two above assumptions, and Figure \ref{fig_darkcore24}a shows the 24 \um\ intensity profiles.  The intensities were obtained by averaging over concentric annuli, each with radii of 2.5\as\ at 8 \um\ and 5\as\ at 24 \um, centered at the position of L673-7-IRS.  The positions plotted on the x-axis show the mid-points of the annuli.  This analysis assumes that the core is spherically symmetric.  The error bars show the standard deviation of the mean for each annulus; note that the actual standard deviations are much larger since the core is not perfectly spherically symmetric (see Figures \ref{fig_3color} and \ref{fig_cores2deeper}).

Inspection of the 8 and 24 \um\ intensity profiles shows that both profiles flatten out at $\sim$70\as, are relatively flat between $\sim 70-100$\as, and then decrease again beyond 100\as.  Thus we define the edge of the core as traced by the absorption profile to be at a projected distance of 70\as\ from the core center (corresponding to 21,000 AU at a distance of 300 pc), and average the values of $I_{\nu}$ between $70-100$\as\ to obtain the background intensities $I_{\nu}^{o}$.

With $I_{\nu}^{o}$ in hand, the optical depth through each annulus, $\tau_{\nu}$, is calculated as

\begin{equation}\label{eq_tau}
\tau_{\nu} = ln\, \left (\frac{I_{\nu}^{o}}{I_{\nu}}\right ) \quad .
\end{equation}
The 8 and 24 \um\ optical depth profiles are shown in Figures \ref{fig_darkcore8}b and \ref{fig_darkcore24}b, respectively.  From the definition of optical depth as $\tau_{\nu} = \int \kappa_{\nu} \rho (r) dr$, the total gas column density in each annulus is then

\begin{equation}\label{eq_N}
N = 100 \, \frac{\tau_{\nu}}{\kappa_{\nu} \mu m_{\rm H}} \quad ,
\end{equation}
where $\kappa_{\nu}$ is the dust opacity, $\mu$ is the mean molecular weight per free particle ($\mu = 2.37$ for a gas that is 71\% by mass hydrogen, 27\% helium, and 2\% metals; Kauffmann et al.~2008), $m_{\rm H}$ is the mass of hydrogen, and the factor of 100 is the assumed gas-to-dust ratio.

For consistency with the dust radiative transfer models in \S \ref{models}, we adopt the dust opacities of Ossenkopf \& Henning (1994) appropriate for thin ice mantles after $10^5$ yr of coagulation at a gas density of $10^6$ cm$^{-3}$ (OH5 dust).  The OH5 dust opacities give $\kappa_{\nu}^{8 \mu \rm m} = 1063.8$ cm$^{2}$ g$^{-1}$ and $\kappa_{\nu}^{24 \mu \rm m} = 852.2$ cm$^{2}$ g$^{-1}$.  For comparison, the Weingartner \& Draine (2001; hereafter WD01) $R_{\rm V} = 5.5$ dust opacities are about a factor of 1.5 lower than the OH5 opacities at 8 and 24 \um.  However, Chapman et al.~(2009a, 2009b) have shown that the mid-infrared extinction law in dense regions flattens considerably more than accounted for by the WD01 $R_{\rm V} = 5.5$ dust model and that the actual opacities may be up to a factor of 2 higher than the WD01 $R_{\rm V} = 5.5$ opacities at these wavelengths.  Thus we adopt the higher OH5 dust opacities but acknowledge there is about a factor of 2 uncertainty in these values.  With the adopted dust opacities the total gas column density in each annulus is calculated from Equation \ref{eq_N} and the results from the 8 and 24 \um\ dark cores are shown in Figures \ref{fig_darkcore8}b and \ref{fig_darkcore24}b, respectively.

The mass in each annulus is then calculated as

\begin{equation}\label{eq_mass}
M = \mu m_{\rm H} N A \quad ,
\end{equation}
where $A$ is the physical area of the annulus calculated assuming a distance of 300 pc.  Figures \ref{fig_darkcore8}c and \ref{fig_darkcore24}c show the total enclosed mass calculated from the 8 and 24 \um\ dark cores, respectively.  Based on the 8 \um\ dark core, the lower and upper limits to the total core mass are 0.6 and 1.9 \msun.  Based on the 24 \um\ dark core, these limits are 0.2 and 4.5 \msun.  These results are summarized in Table \ref{tab_coremass}.

Also plotted on Figures \ref{fig_darkcore8} and \ref{fig_darkcore24} are the column density and total enclosed mass profiles from the best-fit $p=1.5$ and $p=2.0$ dust radiative transfer models.  A description of these models, along with a discussion of how well they match the column density and enclosed mass profiles calculated from the 8 and 24 \um\ dark cores, is presented in \S \ref{models}.

\section{Dust Radiative Transfer Models}\label{models}

In order to obtain an accurate constraint on the internal luminosity of L673-7-IRS, we have constructed physical models of this source, constrained primarily by the source SED (Figure \ref{fig_sed}) but also by the 350 \um\ and 1.2 mm intensity profiles.  The modeling procedure follows closely that presented by Dunham et al.~(2006) (see also Young et al.~2004; Bourke et al.~2006; Lee et al.~2009; Young et al.~2009; Lee et al.~2010).

We used the two-dimensional, axisymmetric, Monte Carlo dust radiative transfer code RADMC (Dullemond \& Turolla 2000; Dullemond \& Dominik 2004) to calculate the temperature profile and emergent SED of a protostar embedded in a dense core (the term envelope is often used interchangeably with dense core in the literature and in this paper).  Despite RADMC allowing for two-dimensional structure, we only consider spherically symmetric models to avoid introducing too many free parameters into models constrained by only a limited number of data points (as given in Table \ref{tab_sed}).  For the dust properties, we adopt the dust opacities of Ossenkopf \& Henning (1994) appropriate for thin ice mantles after $10^5$ yr of coagulation at a gas density of $10^6$ cm$^{-3}$ (OH5 dust), which have been shown to be appropriate for cold, dense cores (e.g., Evans et al.~2001; Shirley et al.~2005).  Isotropic scattering off dust grains is included in the model as described by Dunham et al.~(2010).

The envelope is heated both internally by the protostar and externally by the ISRF.  The external heating is included in our model; we adopt the Black (1994) ISRF, modified in the ultraviolet to reproduce the Draine (1978) ISRF.  Most of the models of low-luminosity, embedded protostars presented in the references listed above then attenuate this Black-Draine ISRF by some $A_V$ of dust with properties given by Draine \& Lee (1984) in order to provide the correct amount of external heating to match the peak of the observed SED.  The physical justification for this attenuation is to simulate the core being embedded in a parent cloud.  For our models of L673-7 any such attenuation causes the model to under-predict the 350 \um\ emission.  Increasing the internal luminosity (see below for more details on the internal source) enables the model to match the observed 350 \um\ emission, but over-predicts the amount of $3.6-70$ \um\ emission.  The $3.6-24$ \um\ model emission can be adjusted by changing the envelope inner radius (again, see below for more details), but the 70 \um\ emission, which primarily traces internal luminosity (Dunham et al.~2008), remains over-predicted.  Thus, we do not attenuate the ISRF.  Ultimately there is a degeneracy between attenuation by surrounding material and intrinsic strength of the ISRF that cannot be broken by SED modeling, particularly an SED as sparsely sampled near the peak as our SED of L673-7.  Observations at $100-500$ \um\ with, for example, the \emph{Herschel Space Observatory}, will be required to better constrain the peak of the SED and total amount of external heating.

With the ISRF constrained as described above, the mass of the envelope is constrained to be $M_{env} \sim 2.25$ \msun\ by requiring that the model flux match the optically thin 1.2 mm continuum flux.  We assume a power-law radial density profile [$n(r) \propto r^{-p}$] for the envelope and, following Young et al.~(2009), we run two model grids, one with $p=1.5$ and one with $p=2.0$.  Estimates of the total core size range from $\sim$10,000 AU from the SHARC-II 350 \um\ data to $\sim$40,000 AU from the MAMBO 1.2 mm data, with the 8 and 24 \um\ absorption profiles suggesting an intermediate value of $\sim$21,000 AU.  We adopt an envelope outer radius of 25,000 AU because it is both the midpoint between the SHARC-II and MAMBO source sizes and close to the core size derived from the absorption profiles, but note that the true core size remains uncertain.  The envelope inner radius ($r_{env}^{in}$) is allowed to vary between $1-300$ AU as 1, 5, 10, 20, 40, 60, ..., 300 AU, for a total of 18 different inner radii.

The internal source consists of a protostar and a circumstellar disk.  The protostar is assumed to be a 3000 K blackbody; as noted by Dunham et al.~(2006), these models are not sensitive to the protostellar temperature since most of the protostellar emission is reprocessed to long wavelengths.  The protostellar luminosity ($L_{star}$) is allowed to vary between $0.0005-0.04$ \lsun\ as 0.0005, 0.001, 0.005, 0.01, 0.015, ..., 0.04 \lsun, for a total of 10 different $L_{star}$.  The presence of a disk is inferred from the presence of a molecular outflow (\S \ref{outflow}) and is in fact required for the model to match the 24 \um\ flux.  Following Dunham et al.~(2006), we include a disk in our 1-D model using the method developed by Butner et al.~(1994), based on the disk model of Adams et al.~(1988).  This method simulates a disk by calculating the emission from a disk with given surface density and temperature profiles at a given inclination, averaging the emission over all inclinations, and then adding this average emission spectrum to the protostellar spectrum to form the final input SED of the internal source used to calculate the radiative transfer through the envelope.  Both the surface density and temperature profiles are described as power laws: $\Sigma(R_{disk}) \propto R_{disk}^{-w}$ and $T(R_{disk}) \propto R_{disk}^{-q}$.  We set  $w$ = 1.5, following Butner et al.~(1994) and Chiang \& Goldreich (1997), and $q$ = 0.5 to simulate a flared disk.  The model also includes an intrinsic disk luminosity, $L_{disk}$, perhaps generated by accretion within or onto the disk.  $L_{disk}$ varies over the same range as above for $L_{star}$.

The inner radius of the disk is set to be the radius at which the temperature is equal to the dust destruction temperature, calculated assuming spherical, blackbody dust grains and a dust destruction temperature of 1500 K.  The outer radius is set to the centrifugal radius.  Given the sound speed, $c_s$, the age of the object, $t$, and the angular velocity of the cloud prior to collapse, $\Omega_{\rm 0}$, the centrifugal radius is calculated as $R_C = (m_0^3/16) c_s t^3 \Omega_{\rm 0}^2$, where $m_0$ is a dimensionless constant of order unity (Terebey et al. 1984; Young \& Evans 2005).  Assuming a gas temperature of 10 K, a sound speed calculated from equal thermal and turbulent contributions, $t=6.1 \times 10^4$ yr (outflow dynamical time; see \S \ref{fco}), and $\Omega_{\rm 0}=4 \times 10^{-13}$ s$^{-1}$ (\andre\ et al.~1999), we calculate $R_C = 128$ AU.  For all models with $r_{env}^{in} > 128$ AU, the disk outer radius is set to $R_C$; otherwise, the disk outer radius is set to $r_{env}^{in}$.  The disk mass is assumed to be $M_{disk} = 0.01$ \msun.  Neither the disk outer radius nor the disk mass are constrained by these models since the long-wavelength emission detected by the MAMBO data is dominated by the envelope.  Sensitive submillimeter and/or millimeter continuum interferometer observations are required to study the disk size and mass (e.g., J\o rgensen et al.~2009; Enoch et al.~2009b).

With the above assumptions, we ran a grid of models over the two possible values of $p$, 18 possible values of $r_{env}^{in}$, 10 possible values of $L_{star}$, and 10 possible values of $L_{disk}$, for a total of 3600 models, 1800 each for the $p=1.5$ and $p=2.0$ density profile power-law indices.  Each model is compared to the observed SED by calculating a reduced chi-squared ($\chi_r^2$) as

\begin{equation}\label{eq_chisquared}
\chi_r^2 = \frac{1}{k}\displaystyle\sum_{i=0}^n \frac{[S_{\nu}^{obs}(\lambda_i)-S_{\nu}^{mod}(\lambda_i)]^2}{\sigma_{\nu}^2(\lambda_i)} \quad , 
\end{equation}
where $S_{\nu}^{obs}(\lambda_i)$ is the observed flux density at $\lambda_i$, $S_{\nu}^{mod}(\lambda_i)$ is the flux density from the model at this wavelength (calculated in an aperture size identical to that used to measure the photometry), and $\sigma_{\nu}(\lambda_i)$ is the uncertainty in the observed flux density.  For $n$ data points and $m$ free parameters there are $k=n-m$ degrees of freedom in these models, and we divide by $k$ to obtain the reduced $\chi^2$ values. There are four data points (excluding the 450 \um\ point) and three free parameters for each of the $p=1.5$ and $p=2.0$ model grids ($r_{env}^{in}$, $L_{star}$, and $L_{disk}$), resulting in one degree of freedom ($k=1$).  Finally, we check whether each model is consistent with the $3.6-8.0$ and 160 \um\ upper limits\footnote{The 160 \um\ upper limit listed in Table \ref{tab_sed} is an upper limit for a point source; an extended source could have a higher total flux and still be undetected.  To compare the models with this upper limit, we calculate the model flux within an aperture with a radius equal to twice the standard deviation of the 160 \um\ PSF and compare this result to the derived upper limit, since nearly all (95\%, to be precise) of the flux from a point-source would be contained within this aperture.  Assuming the PSF is a Gaussian with a FWHM of 40\as, this radius is 34\as.}; only models consistent with these upper limits are kept for further analysis.

\begin{figure}
\plotone{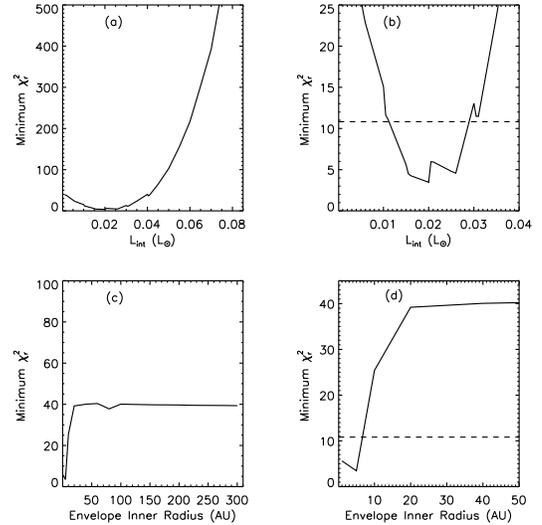}
\caption{\label{fig_chisquared_p15}Results from the $p=1.5$ model grid.  \emph{(a)}: Minimum value of $\chi_r^2$ for each value of \lint\ in the model grid.  \emph{(b)}: Same as (a), except with different x and y axis ranges.  The dotted line shows the 99.9\% confidence limit of 10.827.  \emph{(c)}:  Minimum value of $\chi_r^2$ for each value of $r_{env}^{in}$ in the model grid.  \emph{(d)}: Same as (c), except with different x and y axis ranges.  The dotted line shows the 99.9\% confidence limit of 10.827.}
\end{figure}

Figure \ref{fig_chisquared_p15} shows the results from the $p=1.5$ model grid.  Figures \ref{fig_chisquared_p15}a and \ref{fig_chisquared_p15}b show the minimum value of $\chi_r^2$ for each value of \lint\ (\lint\ $= L_{star}+L_{disk}$) in the model grid.  We do not show results for $L_{star}$ and $L_{disk}$ separately because, as noted by Dunham et al.~(2006), these models are sensitive to the total \lint\ but not particularly sensitive to the individual stellar and disk contributions.  From Figure \ref{fig_chisquared_p15}a it is clear that the best-fit models have \lint\ $\sim$ 0.02 \lsun, and Figure \ref{fig_chisquared_p15}b demonstrates that imposing a 99.9\% confidence limit of $\chi_r^2 \leq 10.827$ constrains $0.01 \leq$ \lint\ (\lsun) $\leq 0.03$.  Figures \ref{fig_chisquared_p15}c and \ref{fig_chisquared_p15}d, which plot the minimum value of $\chi_r^2$ for each value of $r_{env}^{in}$ in the model grid, show that a similar 99.9\% confidence limit constrains $r_{env}^{in} \la$ 5 AU.  This constraint on the inner envelope radius should not be over-interpreted; it is simply the $r_{env}^{in}$ that produces the correct total amount of optical depth through the model to match the observed SED, given the assumed density profile.  Different density profiles would require different $r_{env}^{in}$ to give the same total opacity (see below).

\begin{figure}
\plotone{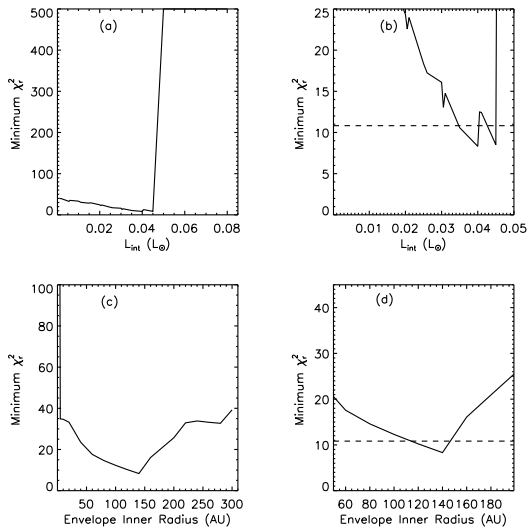}
\caption{\label{fig_chisquared_p20}Same as Figure \ref{fig_chisquared_p15}, except now showing the results from the $p=2.0$ model grid.}
\end{figure}

Figure \ref{fig_chisquared_p20} is similar to Figure \ref{fig_chisquared_p15} but now shows the results for the $p=2.0$ model grid.  The 99.9\% confidence limits constrain $0.035 \leq$ \lint\ (\lsun) $\leq 0.045$ and $100 \leq$ $r_{env}^{in}$ (AU) $\leq 150$.  Again, the constraint on the inner envelope radius should not be over-interpreted.  The constraints on the internal luminosity, however, are robust, as demonstrated by the similar constraints independent of assumed density profile and as discussed by Dunham et al.~(2006).

\begin{figure*}
\plotone{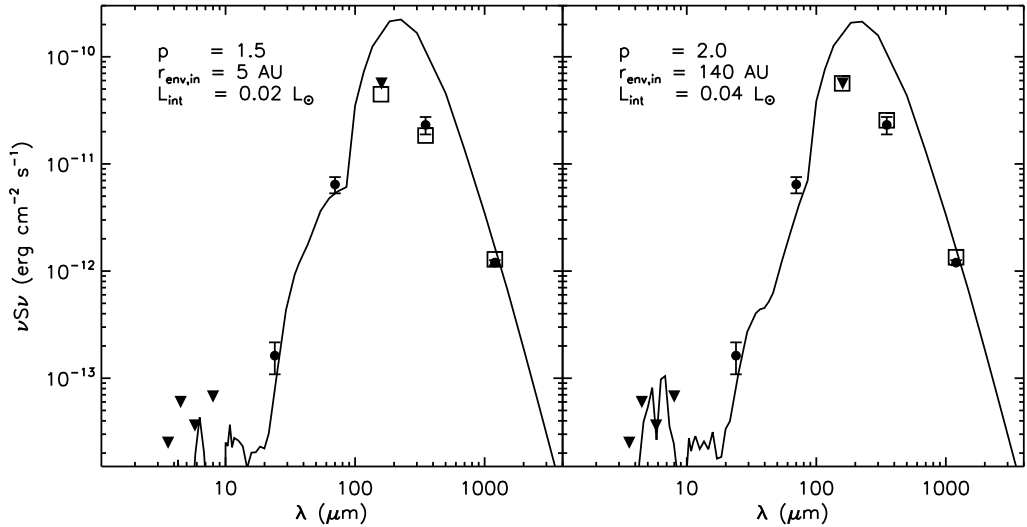}
\caption{\label{fig_modelseds}Best-fit models of L673-7.  The solid lines show the total model fluxes, the filled circles show the photometry listed in Table \ref{tab_sed} (except for the 450 \um\ data point, which is not plotted; see text in \S \ref{sharc} for details), the filled inverted triangles show the upper limits listed in Table \ref{tab_sed}, and the open squares show the model fluxes in the aperture sizes used for photometry at long wavelengths ($\lambda \geq 100$ \um) where the total emission is more extended than the aperture.  \emph{(Left):} Best-fit model from the $p=1.5$ grid, with \lint\ $=0.02$ \lsun\ ($L_{star} = 0.015$ \lsun, $L_{disk} = 0.005$ \lsun), $r_{env}^{in} = 5$ AU, and $\chi_r^2 = 3.5$.  \emph{(Right):} Best-fit model from the $p=2.0$ grid, with \lint\ $=0.04$ \lsun\ ($L_{star} = 0.01$ \lsun, $L_{disk} = 0.03$ \lsun), $r_{env}^{in} = 140$ AU, and $\chi_r^2 = 8.3$.}
\end{figure*}

Figure \ref{fig_modelseds} shows the SED of L673-7 with the best-fit models from the $p=1.5$ and $p=2.0$ grids overplotted.  The best-fit $p=1.5$ model has \lint\ $=0.02$ \lsun\ ($L_{star} = 0.015$ \lsun, $L_{disk} = 0.005$ \lsun), $r_{env}^{in} = 5$ AU, and $\chi_r^2 = 3.5$, while the best-fit $p=2.0$ model has \lint\ $=0.04$ \lsun\ ($L_{star} = 0.01$ \lsun, $L_{disk} = 0.03$ \lsun), $r_{env}^{in} = 140$ AU, and $\chi_r^2 = 8.3$.  Both models are consistent with the $3.6-8.0$ \um\ and 160 \um\ upper limits.  The $p=1.5$ model appears to provide a slightly better fit by eye, a fact that is confirmed by the lower $\chi_r^2$, and also appears more consistent with the upper limits (although, strictly speaking, no upper limits are violated by either model).  In general, however, we are unable to distinguish between these two models based on the SEDs alone.

\begin{figure*}
\plotone{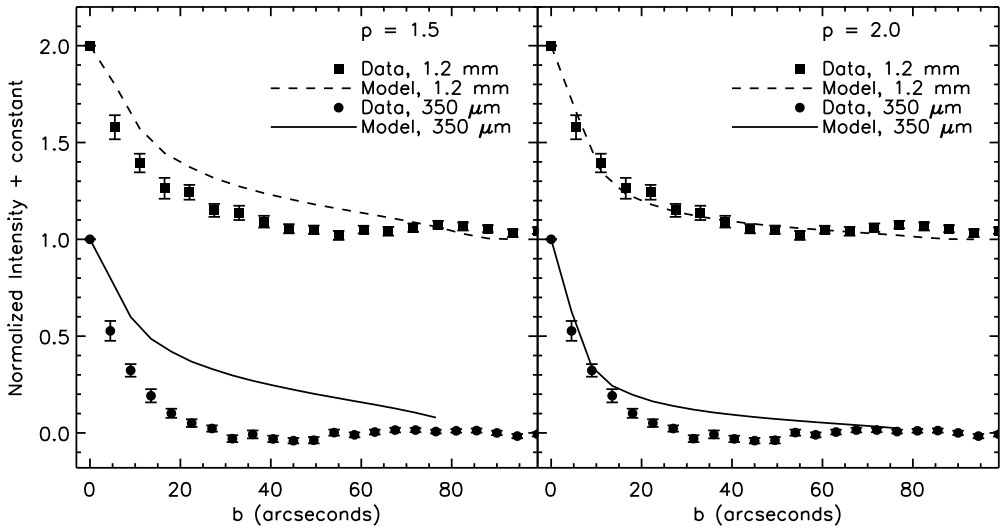}
\caption{\label{fig_intensityprofiles}Intensity profiles for L673-7.  The solid and dashed lines show the 350 \um\ and 1.2 mm model intensity profiles, respectively, for the best-fit models from the $p=1.5$ \emph{(left)} and $p=2.0$ \emph{(right)} grids shows in Figure \ref{fig_modelseds}.  The filled circles and squares show the observed 350 \um\ and 1.2 mm intensity profiles, respectively.}
\end{figure*}

Unlike the SEDs, the submillimeter and millimeter intensity profiles are sensitive to the envelope density profile (e.g., Shirley et al.~2002).  Figure \ref{fig_intensityprofiles} plots the model 350 \um\ and 1.2 mm intensity profiles for the best-fit $p=1.5$ and $p=2.0$ models shown in Figure \ref{fig_modelseds}, along with the observed intensity profiles calculated from the data in a manner essentially identical to the calculation of the 8 and 24 \um\ intensity profiles in \S \ref{darkcore}.  The model intensity profiles for the $p=1.5$ model are clearly too shallow compared to the data.  On the other hand, the $p=2.0$ model intensity profiles provide a much better match to the observed profiles.

Finally, we overplot the column density and enclosed mass profiles calculated from the best-fit $p=1.5$ and $p=2.0$ models on Figures \ref{fig_darkcore8} and \ref{fig_darkcore24}.  We refrain from making any quantitative comparisons between the model and absorption profiles since we were only able to set lower and upper limits to the absorption profiles.  Qualitatively, the shapes of the model profiles generally match those derived from the 8 and 24 \um\ absorption profiles.  The model profile shapes do appear to diverge from the absorption profiles beyond $\sim$ 40\as\ (12000 AU at 300 pc) in that the model column density profiles fall off less steeply and the model enclosed mass profiles become less flat than the absorption profiles.  This divergence can easily be explained by the fact that the models assume an outer envelope radius of 25,000 AU, while the absorption profiles flatten out to the assumed background intensity at 21,000 AU.  These same best-fit models modified to have outer envelope radii of 21,000 AU do indeed provide better matches to the overall shapes of the absorption profiles (not shown).

Based on the above results, we conclude that L673-7 features a relatively steep density profile, with $p \sim 2$.  A more accurate constraint on the exact density profile, particularly at small radii, would require both better limits on the absorption profiles (through better knowledge of the foreground intensities) and quantitatively testing more than just two density profiles, and is beyond the scope of this paper.  From the fits to the SED alone we find that $0.01 \leq$ \lint\ (\lsun) $\leq 0.045$.  Once the constraint on the density profile is added in, we conclude that the most likely value is \lint\ $\sim$ 0.04 \lsun.  This is in very good agreement with the result from Section \ref{evolstatus} that \lint\ $=0.04$ \lsun, based on the Dunham et al.~(2008) relation between \lint\ and 70 \um\ flux.

\section{Outflow Properties and Protostellar Mass Accretion Rate}\label{outflowmdot}

\subsection{Outflow Properties}\label{fco}

We calculate the mass of the outflow, $M_{flow}$, and its kinematic and dynamic properties (kinetic energy, $E_{flow}$, momentum, $P_{flow}$, luminosity, $L_{flow}$, and force\footnote{The outflow force is also commonly referred to as the outflow momentum flux.  It is calculated as a momentum over time and therefore is in fact a force.  Since the usual definition of the flux of some quantity is quantity per area per time, rather than quantity per time, we refrain from calling $F_{flow}$ the momentum flux in this paper.}, $F_{flow}$) in the standard manner (e.g., Cabrit \& Bertout 1990).  To calculate the mass (and all the other properties that depend on mass), we first calculate the column density of H$_2$, $N_{\mathrm{H_2}}$, within each velocity channel in each pixel.  Assuming LTE, $N_{\mathrm{H_2}} = f(T,X_{\rm CO}) (\int \tmb \, \mathrm{d}$$ v)$, where $f(T,X_{\rm CO})$ is a function of the temperature of outflowing gas, $T$, and CO abundance relative to \hh, $X_{\rm CO}$.  The integral $\int \tmb \, \mathrm{d}$$ v$ is over the velocity channel and is given by the temperature in that channel multipled by the channel width (0.13 \kms).  Following Hatchell et al.~(2007), we assume a standard CO abundance relative to \hh\ of $X_{\rm CO} = 10^{-4}$.  This abundance is generally uncertain to within a factor of $\sim 3$ (e.g., Frerking et al.~1982; Lacy et al.~1994), directly leading to a corresponding factor of $\sim 3$ uncertainty in $M_{flow}$ and all other quantities that depend on $M_{flow}$.

The correct temperature to assume is unknown.  Hatchell et al.~(2007) chose 50 K, citing evidence for warm gas in outflows (e.g., Hatchell et al.~1999).  Here we take a different approach.  The temperature is only one of several sources of uncertainty in deriving the outflow properties.  Ultimately, strong lower limits to these properties are more useful than highly uncertain values that attempt to ``average out'' the various sources of uncertainty.  Thus, we choose the temperature from the range of $10-100$ K (encompassing the likely range for gas in molecular outflows) that minimizes $f$, 17.6 K.  Choosing 100 K instead, the temperature that maximizes $f$, would increase the calculated column density by a factor of 2.5.  With $T=17.6$ K, 

\begin{equation}
N_{\mathrm{H_2}} = 5.144 \times 10^{18} \; \left(\int \tmb \, \mathrm{d} v\right) \quad \mathrm{cm}^{-2} \qquad .
\end{equation}
The mass within each velocity channel in each pixel is then calculated as $M_{v, pixel} = \mu_{\rm H_2} m_{\rm H} N_{\mathrm{H_2}} A_{pixel}$, where $m_{\rm H}$ is the mass of a hydrogen atom, $\mu_{\rm H_2}$ is the mean molecular weight per hydrogen molecule ($\mu_{\rm H_2} = 2.8$ for gas composed of 71\% hydrogen, 27\% helium, and 2\% metals by mass [Kauffmann et al.~2008]), and $A_{pixel}$ is the area of each pixel.  The total mass of the flow is then obtained by summing $M_{v, pixel}$ over all velocity channels and all pixels.  Following Hatchell et al.~(2007), we include all velocity channels for which $|v-v_{core}| = 2-15$ \kms\ (where $v_{core}$ is the core systemic velocity).  Since we do not know the total extent of the outflow, we restrict the summation to those pixels within 150\as\ of L673-7-IRS (the approximate extent of the main blue and red lobes; see \S \ref{outflow}) and consider the result a lower limit to the total outflow mass.  Following this procedure, we calculate $M_{flow} \geq 0.05$ \msun\ and list this result in Table \ref{tab_outflow}.

\begin{deluxetable*}{lll}
\tabletypesize{\scriptsize}
\tablewidth{0pt}
\tablecaption{\label{tab_outflow}Outflow and Mass Accretion Properties}  
\tablehead{\colhead{Quantity} & \colhead{Units} & \colhead{Value}}
\startdata
Outflow Mass ($M_{flow}$) & \msun & $\geq$ 0.05 \\
Outflow Kinetic Energy ($E_{flow}$) & $10^{42}$ ergs & $\geq$ 6.1 $\frac{1}{\mathrm{cos}^2 i}$ \\
Outflow Momentum ($P_{flow}$) & \msun\ \kms & $\geq$ 0.13 $\frac{1}{\mathrm{cos} \, i}$ \\
Outflow Luminosity ($L_{flow}$) & $10^{-4}$ \lsun & $\geq$ 8.3 $\frac{\mathrm{sin} \, i}{\mathrm{cos}^3 \, i}$ \\
Outflow Force ($F_{flow}$) & $10^{-6}$ \msun\ \kms\ yr$^{-1}$ & $\geq$ 2.1 $\frac{\mathrm{sin} \, i}{\mathrm{cos}^2 \, i}$ \\
Outflow Force\tablenotemark{a} ($F_{flow}$) & $10^{-6}$ \msun\ \kms\ yr$^{-1}$ & $\geq$ 4.5 $\frac{\mathrm{sin} \, i}{\mathrm{cos}^2 \, i}$ \\
Average Protostellar Mass Accretion Rate ($\langle \dot{M}_{\mathrm acc} \rangle$) & $10^{-6}$ \msun\ yr$^{-1}$ & $\geq$ 1.2 $\frac{\mathrm{sin} \, i}{\mathrm{cos}^2 \, i}$ \\
Dynamical Time ($\tau_{\mathrm d}$) & 10$^4$ yr & $\geq$ 6.1 $\frac{\mathrm{cos} \, i}{\mathrm{sin} \, i}$ \\
Protostellar Mass Accreted ($M_{\rm acc}$) & \msun & $\geq 0.07 \, \frac{1}{\mathrm{cos} \, i}$ \\
Accretion Luminosity\tablenotemark{b} $\,$ (\lacc) & \lsun & $\geq$ 0.36 \lsun \\
\enddata\\
\tablenotetext{a}{Calculated using the force per beam method (Fuller \& Ladd 2002; Hatchell et al.~2007).}
\tablenotetext{b}{Assuming constant mass accretion so that average rate is equal to current rate.}
\end{deluxetable*}

The outflow kinetic energy and momentum within each velocity channel in each pixel are calculated as $E_{v, pixel} = \frac{1}{2} M_{v, pixel} \times v^2$ and $P_{v, pixel} = M_{v, pixel} \times v$, respectively, where $v$ is the velocity of each channel (with respect to the core systemic velocity).  The total kinetic energy and momentum of the outflow are then calculated by summing over all pixels and velocity channels, again including only those channels for which $|v-v_{core}| = 2-15$ \kms\ and pixels within 150\as\ of L673-7-IRS.  We calculate $E_{flow} \geq 6.1 \times 10^{42} \frac{1}{\mathrm{cos}^2 i}$ ergs and $P_{flow} \geq 0.13 \frac{1}{\mathrm{cos} \, i}$ \msun\ \kms, and list these results in Table \ref{tab_outflow}.  The inclination dependence in both quantities arises from the fact that we measure radial velocities rather than true space velocities and is left in the results at this stage.

The outflow luminosity and force are calculated as $L_{flow} = E_{flow} / \tau_d$ and $F_{flow} = P_{flow} / \tau_d$, respectively, where $\tau_d$ is the dynamical time of the outflow.  Assuming a total extent of at least 150\as\ and a characteristic velocity of $v = P_{flow} / M_{flow} = 3.5$ \kms, we calculate $\tau_d \geq 6.1 \times 10^4 \frac{\mathrm{cos} \, i}{\mathrm{sin} \, i}$ yr, $L_{flow} \geq 8.3 \times 10^{-4} \frac{\mathrm{sin} \, i}{\mathrm{cos}^3 \, i}$ \lsun, and $F_{flow} \geq 2.1 \times 10^{-6} \frac{\mathrm{sin} \, i}{\mathrm{cos}^2 \, i}$ \msun\ \kms\ yr$^{-1}$, and list these results in Table \ref{tab_outflow}.  Again, we leave the inclination dependence in the results at this stage.  These lower limits on the mass, luminosity, and force of this outflow are generally consistent with the correlations between these quantities and the luminosity of the driving source presented by Wu et al.~(2004), after extrapolating by eye these correlations (and their disperstions of several orders of magnitude) to the very low luminosity of L673-7-IRS.  However, since we are only able to determine lower limits to the outflow properties, it is also possible that the L673-7 outflow has a higher mass, luminosity, and/or force for the luminosity of its driving source compared to other outflows.

The outflow force is essentially the average rate at which momentum has been injected into the outflow over the region encompassed by the calculation of the outflow momentum and dynamical time (in this case, 150\as, the extent of the primary red and blue lobes).  However, the above calculation simply divides the total momentum by the total dynamical time, and does not take into account the fact that the total time that outflowing gas has spent in the outflow is a function of the velocity of the gas (at the same distance from the protostar, lower velocity gas has been in the outflow for longer than higher velocity gas).  We thus also calculate the force using the outflow force per beam method (Fuller \& Ladd 2002; Hatchell et al.~2007), which calculates the contributions to the total force separately for each velocity channel and then sums over all velocities.  Again including all velocity channels for which $|v-v_{core}| = 2-15$ \kms, we calculate $F_{flow} \geq 4.5 \times 10^{-6}$ $\frac{\mathrm{sin} \, i}{\mathrm{cos}^2 \, i}$ \msun\ \kms\ yr$^{-1}$.  We list this result in Table \ref{tab_outflow}.  The two calculations of the outflow force agree to within a factor of about 2, but we consider this second calculation to be more accurate and use it in the discussion below.

We have not included an opacity correction of $\frac{\tau}{1-\mathrm{e}^{-\tau}}$ in the calculation of the outflow mass (and thus in all other quantities that depend on the outflow mass) to account for the fact that the \cojtwo\ emission is likely optically thick.  This factor is always greater than or equal to 1, and is typically about 3.5 (Cabrit \& Bertout 1992; Bontemps et al.~1996).  This opacity correction, combined with the discussion above about the LTE excitation temperature, implies that our calculated properties are strong lower limits.

\subsection{Protostellar Mass Accretion Rate}\label{mdot}

Molecular outflows such as the one presented here are driven by the transfer of momentum from a jet/wind ejected by the protostellar system to the ambient medium (e.g., Bontemps et al.~1996; Richer et al.~2000; Arce et al.~2007).  Following Bontemps et al.~1996, conservation of momentum requires that

\begin{equation}
F_{flow} = f_{\rm ent} \dot{M}_w V_w \qquad ,
\end{equation}
where $\dot{M}_w$ is the mass-loss rate in the jet/wind, $V_w$ is the velocity of the jet/wind, and $f_{\rm ent}$ is the entrainment efficiency between the jet/wind and the ambient gas.  This expression can be re-written as

\begin{equation}\label{eq_FcoMdotacc}
F_{flow} = f_{\rm ent} \, \frac{\dot{M}_w}{\dot{M}_{\rm acc}} \, V_w  \dot{M}_{\rm acc} \qquad ,
\end{equation}
where \mdotacc\ is the protostellar mass accretion rate.

The rate at which mass is lost in the jet/wind is correlated with the protostellar mass accretion rate (Bontemps et al.~1996 and references therein).  The term $\dot{M}_w / $\mdotacc\ in Equation \ref{eq_FcoMdotacc} can thus be considered a constant, resulting in a direct proportionality between the protostellar mass accretion rate and the outflow force:

\begin{equation}\label{eq_finalmdot}
\dot{M}_{\rm acc} = \frac{1}{f_{\rm ent}} \, \frac{\dot{M}_{\rm acc}}{\dot{M}_w} \frac {1}{V_w} \, F_{flow} \qquad .
\end{equation}

We assume a typical jet/wind velocity of $V_w \sim 150$ \kms (Bontemps et al.~1996).  Models of jet/wind formation predict, on average, $\dot{M}_w / $\mdotacc\ $\sim$ 0.1 (Shu et al.~1994; Pelletier \& Pudritz et al.~1992; Wardle \& Konigl 1993; Bontemps et al.~1996).  The entrainment efficiency is typically $f_{\rm ent} \sim 0.1-0.25$ (Bontemps et al.~1996; \andre\ et al.~1999).  Since we calculated a lower limit to $F_{flow}$ and thus will also calculate a lower limit to \mdotacc, we adopt $f_{\rm ent} = 0.25$.  With these values and our result for $F_{flow}$, we calculate $\langle \dot{M}_{\rm acc} \rangle \geq 1.2 \times 10^{-6} \frac{\mathrm{sin} \, i}{\mathrm{cos}^2 \, i}$ \msun\ yr$^{-1}$, close to the canonical value of $\sim 2 \times 10^{-6}$ \msun\ yr$^{-1}$ for the inside-out collapse of a singular isothermal sphere (Shu, Adams, \& Lizano 1987).  The brackets indicate that this is the \emph{time-averaged} mass accretion rate over the lifetime of the outflow, since $F_{flow}$ is the average rate at which momentum has been injected.

With knowledge about the age of the outflow and the above result for the time-averaged protostellar mass accretion rate, we can calculate the protostellar mass accreted.  For the age of the outflow we use the dynamical time, $\tau_d \geq 6.1 \times 10^4 \frac{\mathrm{cos} \, i}{\mathrm{sin} \, i}$ yr.  The protostellar mass accreted is then $M_{\rm acc} = \dot{M}_{\rm acc} \, \tau_d$.  We calculate $M_{\rm acc} \geq 0.07 \, \frac{1}{\mathrm{cos} \, i}$ \msun\ and list this result in Table \ref{tab_outflow}.

We now have a lower limit to both the time-averaged protostellar mass accretion rate and the protostellar mass.  With these quantities we can calculate the accretion luminosity, 

\begin{equation}\label{eq_lacc}
\lacc = \frac{G M_{\rm acc} \mdotacc}{R} \qquad ,
\end{equation}
where $G$ is the gravitational constant and $R$ is the protostellar radius (assumed to be 3 \rsun\footnote{A larger accretion radius, as would be the case if material were accreting only onto a disk, would lead to a correspondingly smaller accretion luminosity.  Indeed, most material accreting from the envelope likely accretes onto the disk (e.g., Terebey, Shu, \& Cassen 1984; Young \& Evans 2005).  However, the presence of the outflow argues in favor of accretion from the disk onto the protostar, and since $R_{\rm protostar} << R_{\rm disk}$, Equation \ref{eq_lacc} is a good approximation of the accretion luminosity generated by this accretion.}).  With our results from above, we find that \lacc\ $\geq 0.9 \, \frac{\mathrm{sin} \, i}{\mathrm{cos}^3 \, i}$ \lsun.  Through this point we have left the inclination dependence in our results.  Since we are calculating a lower limit to the accretion luminosity, we evaluate the expression $\frac{\mathrm{sin} \, i}{\mathrm{cos}^3 \, i}$ at the inclination that minimizes this quantity.  We restrict ourselves to the range $20\degree \leq i \leq 70\degree$ since extreme pole-on and edge-on inclinations can be ruled out (\S \ref{outflow}).  The minimum value of 0.41 occurs at $i$ = 20\degree, while the maximum value occurs at $i$=70\degree\ and is larger by a factor of 57.  Inserting $i$ = 20\degree\ into the above expression for \lacc, we find

\begin{equation}\label{eq_lacc_result}
\lacc \geq 0.36 \, \lsun \qquad .
\end{equation}
We list this result in Table \ref{tab_outflow}.

Our result for the accretion luminosity is a strong lower limit since $\lacc \propto M_{\rm acc} \mdotacc \propto \mdotacc^2 \taud$ and both \mdotacc\ and \taud\ are lower limits.  Specifically, \mdotacc\ can only increase (typically by a factor of $\sim$ 3.5) when an opacity correction is included to account for the fact that the \cojtwo\ emission is likely optically thick, it can only increase (by up to a factor of 2.5) if the outflowing gas is not at the assumed temperature of 17.6 K, and it can only increase (by up to a factor of 57) if the inclination is larger than 20\degree\ (which it likely is; \S \ref{outflow}).  Additionally, the true dynamical time is almost certainly larger by at least a factor of 2 since the map presented here does not cover the full extent of the outflow.  Thus, we conclude with strong confidence that $\lacc \geq 0.36$ \lsun, with the actual value likely much higher.

\section{Discussion}\label{discussion}

With the detection of an infrared source driving a molecular outflow, L673-7 is clearly a protostellar core.  However, as noted in \S \ref{sec_l6737}, three surveys for infall motions towards starless cores included L673-7 but did not find any evidence for infall in this core (Lee, Myers, \& Tafalla 1999; Lee, Myers, \& Plume 2004; Sohn et al.~2007).  The molecular line observations used to search for infall signatures in these three studies featured beam FWHM ranging from $\sim 25 - 65$\as, corresponding to physical sizes of $7500 - 19500$ AU at the assumed distance of 300 pc.  In the Shu (1977) standard model, a wave of infall propages outward at the sound speed, leading to the definition of an infall radius $r_{\rm inf} = c_s t$, where $c_s$ is the sound speed and $t$ the time since collapse began.  Material within the infall radius is collapsing onto the central protostar whereas material outside the infall radius is still at rest.  Assuming a purely thermal sound speed with 10 K gas, the infall radius reaches radii of $7500 - 19500$ AU at $\sim 190,000 - 488,000$ yr after collapse begins.  These times are a factor of $3-8$ larger than the outflow dynamical time.  If the outflow dynamical time represents the age of the protostar then the collapsing region is smaller than the size of the beam, possibly explaining the nondetection of infall signatures.  In reality the derived outflow dynamical time is only a lower limit since we have not mapped the full extent of the outflow, and material outside the infall radius might already be infalling rather than at rest (e.g., Fatuzzo, Adams, \& Myers 2004).  However, both rotation and outflows can mask the kinematic signatures of inall (Lee, Myers, \& Tafalla 1999 and references therein), thus a nondetection of infall signatures does not necessarily imply a lack of infall.

Based on both a strong correlation between the 70 \um\ flux of a protostar and its internal luminosity (Dunham et al.~2008) and dust radiative transfer models constrained by the observed SED and submillimeter and millimeter radial intensity profiles, L673-7-IRS is a Class 0 protostar with \lint\ $=0.01-0.045$ \lsun, with \lint\ $\sim$ 0.04 \lsun\ the most likely value.  Based on the definition of a VeLLO as a protostar embedded within a dense core with \lint\ $\leq 0.1$ \lsun\ (Di Francesco et al.~2007), L673-7-IRS qualifies as a VeLLO.  It is possibly the lowest luminosity VeLLO yet studied, although such a claim is difficult to prove with any certainty given the factor of $\sim$ 2 uncertainty in \lint\ for all VeLLOs studied to date (Young et al.~2004; Dunham et al.~2006; Bourke et al.~2006; Lee et al.~2009).  Given that the dust in the envelope will reach its new equilibrium temperature very quickly following a change in \lint\ ($<< 100$ yr; Draine \& Anderson 1985; Lee 2007; Dunham et al.~2010), this is a very good measurement of the current \lint.

The properties of the molecular outflow, on the other hand, tell a different story.  From the outflow, we calculate an accretion luminosity of $\lacc \geq 0.36$ \lsun.  This is a very strong lower limit for all of the reasons described in \S \ref{outflowmdot}.  This difference between the above values of \lint\ and \lacc\ by a factor of at least 8 can be resolved by invoking non-steady mass accretion.  The \emph{current} mass accretion rate must be much lower than the \emph{time-averaged} rate over the lifetime of the outflow.

This is not a result unique to this source.  Two other embedded protostars have been studied in the same manner and shown to require lower current mass accretion rates than the time-averaged rates inferred by their outflows: IRAM04191 ($\lacc / \lint \sim 25$; \andre\ et al.~1999; Dunham et al.~2006) and L1251A-IRS3 ($\lacc / \lint \geq 4$; Lee et al.~2010).  There is growing evidence that such evidence for non-steady mass accretion is the norm rather than the exception, although more protostars driving outflows must be studied in this manner to evaluate whether such a result is truly universal.  Such a universal result would support the notion detailed in \S \ref{intro} that accretion is episodic and that most time is spent at low accretion rates.  However, we caution that our results for L673-7 do not prove \emph{episodic} accretion, they merely indicate \emph{non-steady} mass accretion.  Models of collapsing cores that are not singular but instead have flattened density profiles at small radii reminiscent of Bonnor-Ebert spheres (Bonnor 1956; Ebert 1955) feature mass accretion rates that rise at early times and then decline at late times (e.g., Foster \& Chevalier 1993; Henriksen, \andre, \& Bontemps 1997; Vorobyov \& Basu 2005b).  Such models could also explain a lower current accretion rate compared to the time-averaged rate, although a decline in the accretion rate by an amount sufficient to give a current accretion rate nearly a factor of 10 lower than the average rate within the required time ($6.1 \times 10^4$ yr, the dynamical time of the outflow) represents a faster decline than generally predicted by such models.  Future work should investigate this in more detail with higher spatial resolution, better-sampled maps of the L673-7 outflow that allow direct calculations of the accretion rate as a function of time.

L673-7-IRS is classified as a Class 0 protostar based on both the calculated \tbol\ and \lbolsmm\ (\S \ref{evolstatus}).  The physical Stage 0 corresponding to a Class 0 SED is a protostar with greater than 50\% of its total system mass still in the envelope (\andre\ et al.~1993), although the effects of geometry and accretion bursts, if present, can often remove the connection between observed Class and physical Stage (e.g., Robitaille et al.~2006; Dunham et al.~2010).  In this case, however, the connection appears to hold: the outflow properties infer an accreted protostellar mass of $M_{\rm acc} \geq 0.07$ \msun, whereas $M_{env} = 0.2 - 4.5$ \msun.  Thus, $M_{env} / M_{acc} \leq 2.9 - 64.3$; most of the mass of the L673-7 protostellar system appears to still be in the envelope.

While L673-7-IRS is consistent with currently having a substellar mass, it is unlikely to remain as such.  The envelope provides a substantial reservoir of mass that can still accrete onto the protostar.  Assuming a typical star formation efficiency in the dense gas of 30\% (e.g., Alves et al.~2007; Enoch et al.~2008; Evans et al.~2009), the envelope mass range of $0.2 - 4.5$ \msun\ indicates that $0.06-1.35$ \msun\ can still be expected to accrete onto the protostar, as long as the current low mass accretion rate indicated by the low \lint\ is not an indication that the accretion process has somehow terminated prematurely.  Asymmetries in the pattern of accretion onto Class 0 protostars can give them drift velocities relative to their cores that may cause them to drift out of the regions of highest density and prematurely terminate accretion despite the presence of a significant mass reservoir (Stamatellos et al.~2005).  Indeed, Huard et al.~(2006) noted an offset between the position of the protostar L1014-IRS and the density peak of the core in which it is embedded, suggesting a drift velocity between protostar and core comparable to that predicted by Stamatellos et al.~(2005).  While both the SHARC-II 350 \um\ and MAMBO 1.2 mm continuum emission peak slightly southwest of L673-7-IRS (by 2.5 and 5\as, respectively), both offsets are comparable to the pointing uncertainties of the two continuum maps.  There is no clear evidence for an offset between L673-7-IRS and its core in the current data, although this question should be revisited with higher-resolution millimeter continuum and extinction maps.

As discussed in \S \ref{sec_l6737}, the distance to L673-7 is quite uncertain.  Herbig \& Jones (1983) concluded that the most likely distance is 300 pc, and we have adopted their result in this paper.  However, as their discussion indicates, the distance could be up to $2-3$ times larger.  The value of \lint, determined either by the 70 \um\ flux or by the radiative transfer models, will essentially just scale as $\lint \propto D^2$.  A larger distance by factors of $2-3$ would thus lead to a larger \lint\ by factors of $4-9$.  L673-7-IRS would no longer be classified as a VeLLO, but, with $\lint \sim 0.16-0.36$ \lsun, it would still be a low-luminosity protostar.  More importantly, the discrepancy between \lint\ and \lacc\ would remain, and in fact would increase.  Since $M_{acc} = \dot{M}_{\rm acc} \tau_d$, $\dot{M}_{\rm acc} \propto F_{flow} \propto D$, and $\tau_d \propto D$, Equation \ref{eq_lacc} gives $L_{acc} \propto \dot{M}_{\rm acc} M_{acc} \propto \dot{M}_{\rm acc}^2 \tau_d \propto D^3$.  For distances $2-3$ times greater than the assumed distance of 300 pc, \lacc\ increases by factors of $8-27$, increasing the discrepancy between \lacc\ and \lint\ by factors of $2-3$.  While the actual values of \lint\ and \lacc\ are sensitive to the assumed distance, the conclusion of this paper that the \emph{current} mass accretion rate must be much lower than the \emph{time-averaged} rate over the lifetime of the outflow remains unchanged even if the distance is much greater than the assumed value of 300 pc.

\section{Conclusions}\label{conclusions}

We have presented new infrared, submillimeter, and millimeter observations of the dense core L673-7 and have discovered a low-luminosity, embedded protostar driving a molecular outflow.  We summarize our main conclusions as:

\begin{enumerate}
\item L673-7 is seen in absorption against the mid-infrared background in 5.8, 8, and 24 \um\ \emph{Spitzer} images, allowing for a derivation of the column density profile and enclosed mass of the core independent of the millimeter continuum data and thus independent of dust temperature.  Estimates of the core mass from the 8 and 24 \um\ absorption profiles range from $0.2-4.5$ \msun.  The MAMBO 1.2 mm emission gives a mass of $\sim$ 2 \msun, both from a direct calculation assuming isothermal dust and from the dust radiative transfer models constrained by the MAMBO data point.  These mass estimates are consistent with each other given the uncertainties in the dust opacities at 8, 24, and 1200 \um\ and the fact that only upper and lower limits to the true mid-infrared absorption profiles could be derived.
\item L673-7-IRS is a Class 0 protostar with \lint\ $= 0.01-0.045$ \lsun, with \lint\ $\sim$ 0.04 \lsun\ the most likely value.  It is thus classified as a VeLLO, and is among the lowest luminosity VeLLOs yet studied.
\item The properties of the molecular outflow suggest a protostellar mass accretion rate, averaged over the lifetime of the outflow, of $\langle \dot{M}_{\mathrm acc} \rangle \geq 1.2 \times 10^{-6} \frac{\mathrm{sin} \, i}{\mathrm{cos}^2 \, i}$ \msun\ yr$^{-1}$, close to the canonical value of $\sim 2 \times 10^{-6}$ \msun\ yr$^{-1}$ for the inside-out collapse of a singular isothermal sphere (Shu, Adams, \& Lizano 1987).  Accretion at this rate over the dynamical time of the outflow indicates that the protostar has grown to at least 0.07 \msun\ in mass.
\item The resulting accretion luminosity from accretion at this rate onto a protostar with this mass is $\lacc \geq 0.36$ \lsun.  The discrepancy between \lint\ and \lacc\ suggests that the current accretion rate is much lower than the time-averaged rate over the lifetime of the outflow, indicating non-steady mass accretion.  To date, two other Class 0 protostars have been shown to have similar mismatches between \lacc\ and \lint.
\item Although L673-7-IRS is consistent with currently being substellar, there is a substantial mass reservoir ($0.2-4.5$ \msun) remaining in the envelope for further accretion, making it highly unlikely that L673-7-IRS will remain substellar.  The high envelope mass relative to the accreted protostellar mass ($M_{env} / M_{acc} \leq 2.9 - 64.3$) indicates that L673-7-IRS is not only a Class 0 protostar based on its SED, but also a Stage 0 protostar with greater than 50\% of its total system mass still in the envelope.
\end{enumerate}

While the results for \lint\ should be robust to additional data (assuming the distance is correct), and the discrepancy between \lacc\ and \lint\ should be robust to both additional data and larger possible distances, future observational studies of L673-7 are needed on several fronts:  (1) A larger area should be mapped in CO in order to discern the true extent of the outflow.  (2) Higher-resolution interferometer CO maps are required to study the outflow on smaller spatial scales close to the protostar, probing the more recent mass accretion history.  (3) Submillimeter/millimeter continuum interferometer data are needed in order to probe basic disk properties (mass, size, etc.).  (4) Better photometry near the peak of the SED ($\sim 100-300$ \um) is needed to constrain the bolometric luminosity and perform a more accurate separation between internal and external heating.  (5) Infrared and (sub)millimeter spectroscopy (with, e.g., \emph{Herschel}, \emph{JWST}, and ground-based facilities) are needed to study both the gas and solid-phase chemistry and search for indications of higher past luminosity (and thus accretion) in the abundances of various species (e.g., Lee 2007).  Once such studies are available, the simple, spherically symmetric models presented in this paper should be revisited in order to place better constraints on source properties, in particular the physical structure of the core.

\acknowledgements
This work is based partly on observations obtained with the \emph{Spitzer Space Telescope}, operated by the Jet Propulsion Laboratory, California Institute of Technology, and the Caltech Submillimeter Observatory.  The authors gratefully acknowledge the assistance of Miranda Dunham in obtaining the CSO \cojtwo\ data, and that of Miwa Block of the MIPS instrument team at Steward Observatory with data reduction.  We thank Cornelis Dullemond for reading a draft of this work and providing helpful comments, Jes J\o rgensen for his IDL scripts to display three-color images, and the Lorentz Center in Leiden for hosting several meetings that contributed to this paper.  This research has made use of NASA's Astrophysics Data System (ADS) Abstract Service and of the SIMBAD database, operated at CDS, Strasbourg, France.  Support for this work, part of the \emph{Spitzer} Legacy Science Program, was provided by NASA through contracts 1224608, 1288664, 1288658, and RSA 1377304.  Support was also provided by NASA Origins grant NNX 07-AJ72G and NSF grant AST-0607793.  MMD acknowledges partial support from a UT Austin University Continuing Fellowship.  Partial support for TLB was provided by NASA through contracts 1279198 and 1288806 issued by the Jet Propulsion Laboratory, California Institute of Technology, to the Smithsonian Astronomical Observatory.


\end{document}